\documentclass[nofootinbib]{revtex4}
\usepackage{graphicx}
\usepackage{dcolumn}
\usepackage{amsmath,amssymb,epsfig}
\usepackage{paralist}
\usepackage{comment}
\usepackage{graphicx}
\usepackage{multirow}
\usepackage{color,soul}
\usepackage{suffix}
\usepackage{mathtools}

\renewcommand{\vec}[1]{\boldsymbol{\mathrm{#1}}}

\begin{document}

\title{Multipole decomposition of gravitational lensing}

\author{Slava G. Turyshev$^{1}$, Viktor T. Toth$^2$}

\affiliation{\vskip 3pt
$^1$Jet Propulsion Laboratory, California Institute of Technology,\\
4800 Oak Grove Drive, Pasadena, CA 91109-0899, USA}

\affiliation{\vskip 3pt
$^2$Ottawa, Ontario K1N 9H5, Canada}

\date{\today}

\begin{abstract}

We study gravitational lensing by a generic extended mass distribution.
For that, we consider the diffraction of electromagnetic (EM) waves by an extended, weakly aspherical, gravitating object. We account for the static gravitational field of such a lens by representing its exterior potential in the most generic form, expressed via an infinite set of symmetric trace free (STF) tensor multipole mass moments. This yields the most general form of the gravitational phase shift, which allows for a comprehensive description of the optical properties of a generic gravitational lens. We found that at each order of the STF moments, the gravitational phase shift is  characterized by only two parameters: a magnitude and a rotation angle that characterize the corresponding caustics, which form in the point spread function (PSF) of the lens. Both of these parameters are uniquely expressed in terms of the transverse-trace free (TT) projections of the multipole moments on the lens plane.  Not only does this result  simplify the development of physically consistent models of realistic lenses, it also drastically reduces the number of required parameters in the ultimate model. To help with the  interpretation of the results, we established the correspondence of the gravitational phase shift expressed via the TT-projected STF multipole mass moments and its representation via spherical harmonics. For axisymmetric mass distributions, the new results are consistent with those that we obtained in previous studies. For arbitrary mass distributions, our results are novel and offer new insight into gravitational lensing by realistic astrophysical systems. These findings are discussed in the context of ongoing astrophysical gravitational lensing investigations as well as observations that are planned with the solar gravitational lens (SGL).

\end{abstract}


\maketitle

\section{Introduction}

Efforts to describe gravitational lensing are often limited to spherically symmetric gravitational fields, where the source of the field is a pointlike, structureless monopole. While this approximation works well in cases of lensing that involve compact lenses and large impact parameters, realistic astrophysical lenses are extended objects with complex structures that are not well approximated by a point-mass representation. Despite the importance of going beyond the point-mass approximation (or that of a Schwarzschild lens), attempts to describe extended gravitational lenses are few in number. They often utilize the geometric optics approximation, which yields divergent light amplification at the caustics of the lens. The presence of caustics, which is a distinguishing feature of extended lenses, requires a wave-optical treatment. Until recently, such a treatment was not readily available in the literature.

To address this need, while investigating the solar gravitational lens (SGL) \cite{Turyshev-Toth:2017},  we developed an approach to study the optical properties of a generic extended gravitational lens \cite{Turyshev-Toth:2021-multipoles}. We considered the propagation of a high-frequency electromagnetic (EM) wave in the vicinity of an extended gravitating body. Using the Mie theory, together with the eikonal approximation, we solved the Maxwell equations on the background of a static gravitational field, while working within the first post-Newtonian approximation of the general theory of relativity. We developed a new approach, called the angular eikonal method, that allows us to study the diffraction pattern formed in the image plane by EM radiation that passed through the gravitational field in proximity to an extended gravitating object.

Our solution is valid for a generic gravitational field. The field is characterized by an infinite set of multipole moments. Aiming at the potential practical applications of the SGL for resolved imaging of faint distant objects (such as distant exoplanets), we considered, in particular, axisymmetric gravitational fields, such as the gravitational field of the Sun.  In this case, the Newtonian gravitational potential is characterized by an infinite set of zonal harmonics. We observed how these harmonics contribute to the diffraction pattern of the EM field, modifying the optical properties of the lens. The presence of gravitational multipoles results in the formation of caustics in the point-spread function (PSF) \cite{Turyshev-Toth:2021-multipoles,Turyshev-Toth:2021-caustics}. It also affects images formed in the focal plane of an imaging telescope \cite{Turyshev-Toth:2021-imaging}.

\begin{figure}
\includegraphics[scale=0.27]{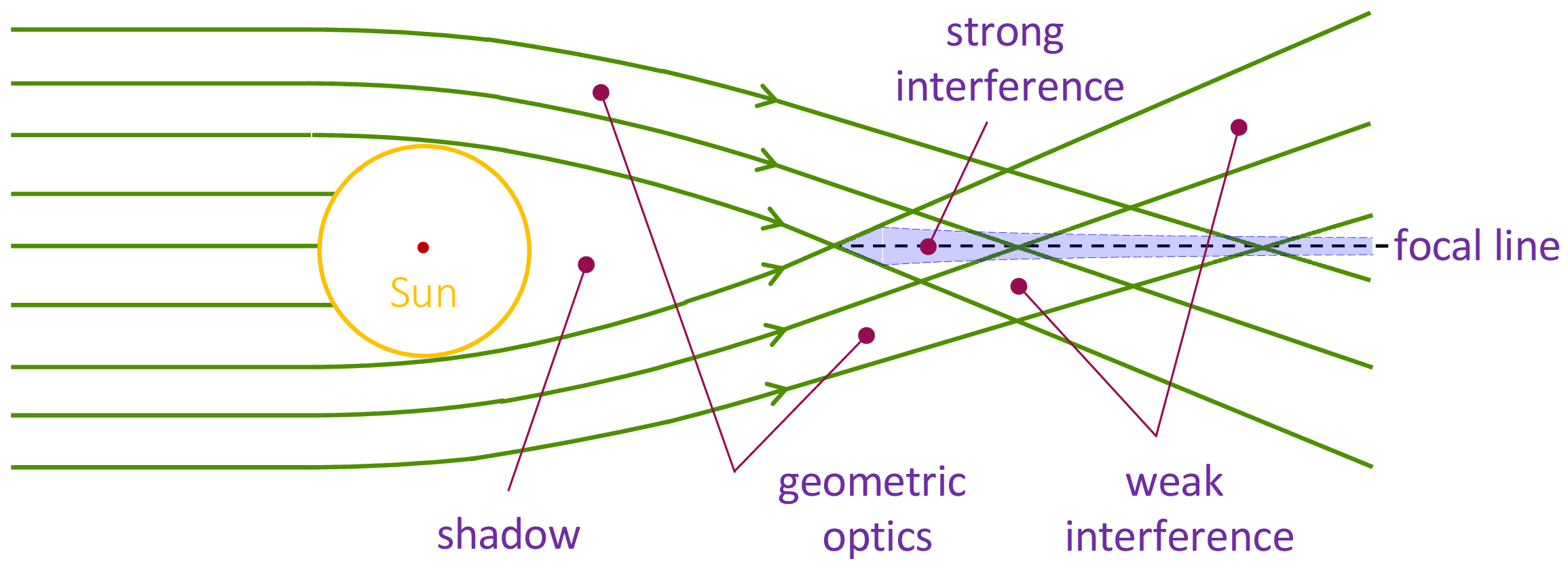}
\caption{\label{fig:regions}The different optical regions of the SGL
(adapted from \cite{Turyshev-Toth:2019-extend}).
}
\end{figure}

In Refs.~\cite{Turyshev-Toth:2021-multipoles,Turyshev-Toth:2021-all-regions} we found the solution to describe the EM field in the image plane, divided into various regions of interest interest, including the regions of strong and weak interference and the region of geometric optics. For that, we studied an optically opaque lens with an arbitrary axially symmetric mass distribution, such as a large rotating star or distant, compact spiral and elliptical galaxies. We have shown that deviations from spherical symmetry in the gravitational field of the lens are relevant only in the strong interference region of the lens in close vicinity to its primary optical axis (a line connecting the center of the target and that of the lens) (see Fig.~\ref{fig:regions}). Such asphericity leads to the appearance of caustics of various order in the PSF of the lens. In the two remaining regions, the optical properties of the lens are consistent with those of a point mass \cite{Turyshev-Toth:2017,Turyshev-Toth:2019-extend,Turyshev-Toth:2020-image,Turyshev-Toth:2020-extend,Toth-Turyshev:2020}.  Hence, to capture the most interesting lensing behavior by a generic lens, we focus on the strong interference region.

As we have shown in \cite{Turyshev-Toth:2021-multipoles} (see Appendix B therein), the new angular eikonal method can be extended to describe lensing by a generic extended lens, with a gravitational field that is sourced by a mass distribution that is given in the form of symmetric trace-free (STF) tensor moments. Representing the gravitational potential of an extended gravitating body in terms of STF tensors is equivalent to the spherical harmonics representation. The advantage of using STF moments is that they allow us to derive the gravitational phase shift in the generic case for arbitrary mass distributions, thus generalizing our previous results for axisymmetric bodies. Recognizing the value of such a major modeling improvement, there is strong motivation to further develop this approach and demonstrate its value for practical applications. This is the purpose of our present paper.

This paper organized as follows:
In Section \ref{sec:opt-prop} we summarize the solution for the EM field that was obtained on the background of a gravitational field with a generic mass distribution. We also discuss the angular eikonal method that can be used to study the optical properties of a lens.
In Section \ref{sec:arb-STF-mom} we consider lensing by bodies of arbitrary composition, with their static gravitational fields represented by STF tensor mass multipole moments. We derive the eikonal gravitational phase shift and discuss the optical properties of a generic gravitational lens. We compute results for several low order moments.  In Section~\ref{sec:generic} we generalize the results to describe gravitational lensing using the entire infinite set of the STF multipole moments.
In Section~\ref{sec:end} we discuss results and outline the next steps in our investigation.
To streamline the main text, we moved some computational details to the Appendices.
Appendix~\ref{sec:stf-comb} presents the lowest order STF moments. The correspondence between STF moments and spherical harmonics is shown in Appendix~\ref{sec:stf-sph-harm}.  In Appendix~\ref{sec:cases} we compute specific derivatives with respect to the vector impact parameter. Appendix~\ref{sec:projops} introduces the projection operators. Polarization matrices of the corresponding order are shown in Appendix~\ref{sec:pol-mat}. Finally, the situation of light propagating at large impact parameter with respect to a quadrupolar lens is briefly explored in Appendix~\ref{sec:beyond}.

\section{Optical properties of an extended lens}
\label{sec:opt-prop}

We consider an isolated extended mass distribution, acting as a gravitational lens. To characterize the gravitational field of a generic lens, following \cite{Turyshev-Toth:2017,Turyshev-Toth:2021-multipoles},  we use a static harmonic metric in the first post-Newtonian approximation of the general theory of relativity.  The line element for this metric in lens-centric spherical coordinates $(r,\theta,\phi)$, to the accuracy sufficient to describe light propagation in a weak gravitational field \cite{Turyshev-Toth:2013}, may be given as
\begin{eqnarray}
ds^2&=&\Big(1+c^{-2}U+{\cal O}(c^{-4})\Big)^{-2}c^2dt^2-\Big(1+c^{-2}U+{\cal O}(c^{-4})\Big)^2\big(dr^2+r^2\big(d\theta^2+\sin^2\theta d\phi^2\big)\big),
\label{eq:metric-gen}
\end{eqnarray}
where the Newtonian potential, $U$, generated by the mass density $\rho({\vec r})$ characterizing the source, is  given as usual:
\begin{eqnarray}
U({\vec r})=G\int\frac{\rho({\vec r}')d^3{\vec r}'}{|{\vec r}-{\vec r}'|}.
\label{eq:w-PN}
\end{eqnarray}

We study the propagation of a high-frequency plane EM wave (i.e., neglecting terms $\propto(kr)^{-1}$, where $k=2\pi/\lambda$ is the wavenumber and $\lambda$ is the wavelength) in the vicinity of the lens.  The high-frequency approximation was justified and used in \cite{Turyshev-Toth:2017,Turyshev-Toth:2020-extend,Turyshev-Toth:2021-multipoles} to allow treatment of the Maxwell equations on the background of (\ref{eq:metric-gen})--(\ref{eq:w-PN}). Defining the object's Schwarzschild radius as $r_g=2GM/c^2$, where $M$ is the object's mass with $\lambda/r_g\ll1$, and $R$ is the object's radius with $r_g/R\ll1$, we assume that the wave is emitted by a point source that is located at a large distance $r_0$ from the lens, so $r_g/r_0\ll 1$, and it is received in an image plane also located at a large distance $r$ from the lens, such that $r_g/r\ll1$.

\subsection{Notations and lensing geometry}

Before summarizing the solution derived in \cite{Turyshev-Toth:2021-multipoles}, we need to present the geometry of the problem and introduce our basic notations. Following \cite{Turyshev-Toth:2021-multipoles,Turyshev-Toth:2017}, we represent the incident (unperturbed) trajectory of a light ray as
{}
\begin{eqnarray}
\vec{r}(t)&=&\vec{r}_{0}+\vec{k}c(t-t_0)+{\cal O}(r_g),
\label{eq:x-Newt0}
\end{eqnarray}
where $\vec k$ is a unit vector in the incident direction of the light ray's propagation path and $\vec r_0$ represents the starting point. Next, we define ${\vec b}=[[{\vec k}\times{\vec r}_0]\times{\vec k}]$ as the impact parameter of the unperturbed trajectory of the light ray.

We introduce the parameter $\tau=\tau(t)$ along the path of the light ray (see details in Appendix~B in \cite{Turyshev-Toth:2017}):
{}
\begin{eqnarray}
\tau &=&({\vec k}\cdot {\vec r})=({\vec k}\cdot {\vec r}_{0})+c(t-t_0),
\label{eq:x-Newt*=0}
\end{eqnarray}
which may be positive or negative. Note that $\tau=z$ when the $z$-axis of the chosen Cartesian coordinate system is oriented along the incident direction of the light ray. We can see that the quantity $\tau=r\cos\alpha$ evolves from a negative value (representing a source at a large distance from the lens, $\alpha \simeq\pi$), through $\tau=0$ (the shortest distance from the lens where $\alpha =\pi/2$), to positive values (with $\alpha \simeq 0$ at the image plane.)
The parameter $\tau$ allows us to rewrite (\ref{eq:x-Newt0}) as
{}
\begin{eqnarray}
{\vec r}(\tau)&=&{\vec b}+{\vec k} \tau+{\cal O}(r_g),
\qquad {\rm with} \qquad ||{\vec r}(\tau)|| \equiv r(\tau) =\sqrt{b^2+\tau^2}+{\cal O}(r_g).
\label{eq:b0}
\end{eqnarray}

We use a lens-centric cylindrical coordinate system $(\rho,\phi,z)$ with its $z$-coordinate oriented along the wavevector $\vec k$, a unit vector in the unperturbed direction of the propagation of the incident wave. We also introduce a light ray's impact parameter, $\vec b$, and coordinates in the image plane, $\vec x$, located in the strong interference region at distance $z$ from the lens.  These quantities are given as:
{}
\begin{eqnarray}
\vec k&=&(0,0,1),
\label{eq:note-k}\\
{\vec b}&=&b(\cos\phi_\xi,\sin \phi_\xi,0)=b\, \vec m,
\label{eq:note-b}\\
{\vec x}&=&\rho(\cos\phi,\sin \phi,0)=\rho \, \vec n.
\label{eq:note-x}
\end{eqnarray}

With these definitions, we may now proceed with summarizing the  solution derived in \cite{Turyshev-Toth:2021-multipoles}.

\subsection{The EM field on the image plane}

As the EM wave travels through a gravitational field, interaction with gravity causes the wave to scatter and diffract \cite{Turyshev-Toth:2017}.  In Ref.~\cite{Turyshev-Toth:2021-multipoles,Turyshev-Toth:2021-all-regions}, while studying the Maxwell equations on the background space-time (\ref{eq:metric-gen}), we developed a solution to the Mie problem for the diffraction of the EM waves on a large gravitating body ($\lambda/R\ll1$, see discussion in \cite{Born-Wolf:1999,Turyshev-Toth:2017})
and found the EM field at an image plane located in any of the optical regions behind the lens (see Fig.~\ref{fig:regions}).

In the cylindrical coordinate system and overall lensing geometry presented above, within the paraxial approximation, this EM field on an image plane takes the following form  \cite{Turyshev-Toth:2021-multipoles,Turyshev-Toth:2021-all-regions}:
{}
\begin{eqnarray}
    \left( \begin{aligned}
{E}_\rho& \\
{H}_\rho& \\
  \end{aligned} \right) =\left( \begin{aligned}
{H}_\phi& \\
-{E}_\phi& \\
  \end{aligned} \right) &=&
\frac{E_0}{r_0}
A(\vec x)e^{-i\omega t}
 \left( \begin{aligned}
 \cos\phi& \\
 \sin\phi& \\
  \end{aligned} \right)+{\cal O}\Big(r_g^2,\rho^2/z^2\Big),
  \label{eq:DB-sol-rho}
\end{eqnarray}
where  $E_0$ is the constant amplitude of the field. The remaining components are small,
 $({E}_z, {H}_z)\propto {\cal O}({\rho}/{z})$.

In its general form, the complex amplitude of the EM field $A(\vec x)$  from (\ref{eq:DB-sol-rho}) is given as
{}
\begin{eqnarray}
A(\vec x) &=&
 \frac{k}{ir}\frac{1}{2\pi}\iint d^2\vec b \,\exp\Big[i\varphi({\vec b},{\vec x})\Big],
  \label{eq:amp-AS}
\end{eqnarray}
where $\varphi$ is the eikonal phase of the EM wave accumulated as the wave travels on its path from the source to the image plane. The eikonal phase is a scalar function
 invariant under a set of general coordinate transformations. Within the required approximation, the phase $\varphi$ is found as a solution to the eikonal equation \cite{Fock-book:1959,Landau-Lifshitz:1988}:
{}
\begin{eqnarray}
g^{mn}\partial_m\varphi\partial_n\varphi&=&0,
  \label{eq:amp-eik}
\end{eqnarray}
which is a direct consequence of Maxwell's equations. Solution of (\ref{eq:amp-eik}) describes the wavefront of an EM wave
propagating in curved spacetime. The solution's geometric
properties are defined by the metric tensor $g_{mn}$ from (\ref{eq:metric-gen}).

A solution to (\ref{eq:amp-eik}), is sought by expanding the eikonal $\varphi$ with respect to the gravitational constant $G$ while assuming that the unperturbed solution is a plane wave (see, for instance, \cite{Turyshev-GRACE-FO:2014} and references therein). Using the parametrization (\ref{eq:note-k})--(\ref{eq:note-x}), the result is given as
{}
\begin{eqnarray}
\varphi(\vec x) &=&
 \varphi_0+
\frac{k}{2 \tilde r}({\vec b} - \vec x)^2+\varphi(\vec b)+{\cal O}\big(r_g^2\big),
  \label{eq:amp-eik-sol}
\end{eqnarray}
where the first term, $\varphi_0=k(r_0+r+{\vec x}^2/2(r_0+r))$, is the phase accumulated by the EM if it were traveled in the absence of gravity with  $1/\tilde r=1/r+1/r_0$ (as discussed in \cite{Turyshev-Toth:2019-extend}). The second term, $k({\vec b} - \vec x)^2/{2 \tilde r}$, is the extra geometric path resulted from gravitational lensing,  treated within the thin lens approximation \cite{Schneider-Ehlers-Falco:1992}.

The last quantity in the phase of this expression is the  gravitational phase shift, $\varphi(\vec b) $, that is acquired by the EM wave as it propagates along its geodetic path from the source to the image plane on the background of the gravitational field (\ref{eq:metric-gen})--(\ref{eq:w-PN}) with potential, $U$, from (\ref{eq:w-PN}) that has the form (see discussion in \cite{Turyshev-Toth:2021-multipoles}):
{}
\begin{eqnarray}
\varphi(\vec b) &=&\frac{2k}{c^2}\int^{\tau}_{\tau_0}  U({\vec b},\tau') d\tau'.
\label{eq:Psi+}
\end{eqnarray}
Note that dependence on $\tau$ in the Newtonian potential comes from the fact that $\vec r\equiv \vec r(\vec b,\tau)$, as given by (\ref{eq:b0}).

Substituting these results in (\ref{eq:amp-AS}) yields the amplification factor of the EM field in the image plane, $A(\vec x)$, which exhibits the familiar structure of the Fermat potential \cite{Turyshev-Toth:2021-multipoles}:
{}
\begin{eqnarray}
A(\vec x) &=&
 \frac{k}{ir}\frac{1}{2\pi}\iint d^2\vec b \,\exp\Big[ik\Big(\frac{1}{2 \tilde r}({\vec b} - \vec x)^2+\frac{2}{c^2}\int^{\tau}_{\tau_0}  U({\vec b},\tau') d\tau'\Big)\Big].
  \label{eq:amp-A}
\end{eqnarray}

The integral of (\ref{eq:amp-A})  is well known. Within the weak field and slow motion approximation, it was obtained previously using different methods and tools. For instance, a similar integral formula for the lensed wave amplitude was obtained using the scalar theory of light in \cite{Deguchi-Watson:1987,Nambu:2013,Nambu:2013a,Matsunaga-Yamamoto:2006}; by using the Fresnel--Kirchhoff diffraction formula \cite{Born-Wolf:1999}; and it was also obtained using the path integral formalism \cite{Feynman:1948,Feynman-Hibbs:1991} in \cite{Nakamura-Deguchi:1999,Yamamoto:2017}.  We note that due to the spherical symmetry of the monopole gravitational field and its conceptual simplicity, the majority of previous efforts considered primarily the case a point mass, where the gravitational potential (\ref{eq:w-PN}) is given only by the monopole term. Only a handful of authors considered  contributions from a quadrupole mass distribution (see relevant discussion in \cite{Turyshev-Toth:2021-multipoles,Turyshev-Toth:2021-caustics}).

Our solution (\ref{eq:DB-sol-rho}) with $A(\vec x)$ from (\ref{eq:amp-A}) generalizes previous results by  treating the EM field as a genuine vector field. It was presented in  \cite{Turyshev-Toth:2021-multipoles} and allows us to develop a wave-optical treatment of gravitational lensing, which is important for practical applications of this phenomenon \cite{Turyshev-etal:2020-PhaseII}, relying on conventional tools of wave optics, e.g., \cite{Goodman:2017}.

\subsection{Optical properties of an axisymmetric lens}
\label{sec:eik-phase-axsym}

To explore the solution (\ref{eq:DB-sol-rho})--(\ref{eq:amp-A}), in Refs.~\cite{Turyshev-Toth:2021-multipoles,Turyshev-Toth:2021-all-regions}, we considered the case of a lens with arbitrary multipole structure. We used a form of $U$ representing an axisymmetric gravitational field of a body (such as the Sun) with mass $M$ and equatorial radius $R$, with its external gravitational potential reduced to $k=0$ spherical harmonics, $C_{\ell0}$, and expressed \cite{Roxburgh:2001,LePoncinLafitte:2007tx} in terms of the usual dimensionless zonal harmonic coefficients $J_\ell=-C_{\ell0}$:
{}
\begin{eqnarray}
U(\vec r)&=&\frac{GM}{r}\Big\{1-\sum_{\ell=2}^\infty J_\ell \Big(\frac{R}{r}\Big)^\ell P_\ell\Big(\frac{{\vec k}\cdot{\vec r}}{r}\Big)\Big\}+
{\cal O}(c^{-4})
=
GM\Big\{\frac{1}{r}-\sum_{\ell=2}^\infty \frac{(-1)^\ell}{ \ell!} J_\ell  R^\ell\frac{\partial^\ell}{\partial s^\ell}\Big(\frac{1}{r}\Big)\Big\}+
{\cal O}(c^{-4}),
\label{eq:pot_stis}
\end{eqnarray}
where $P_\ell$ are the Legendre polynomials \cite{Abramovitz-Stegun:1965}, and ${\vec s}$ is the unit vector along the axis of rotation. Furthermore, in the case of an axisymmetric gravitational field that also exhibits ``north-south symmetry'', such as the Sun's, the expression (\ref{eq:pot_stis}) contains only the $\ell=2,4,6,8...$ even zonal harmonics \cite{Roxburgh:2001}.

To derive the gravitational phase shift $\varphi(\vec b)$ present in (\ref{eq:amp-A}) (and explicitly shown by (\ref{eq:Psi+})), we consider the fact that the typical distances travelled by the EM wave from the source to the lens, $\tau_0=(\vec k\cdot{\vec r}_0)$, and from the lens to the observer, $\tau=(\vec k\cdot{\vec r})$, are much larger than the impact parameter, namely $b/|\tau_0|\ll1$ and $b/|\tau|\ll1$.  Following \cite{Turyshev-Toth:2021-multipoles},  we introduce a unit vector in the direction of the axial symmetry (i.e.,  rotation axis), $\vec s$, given as:
{}
\begin{eqnarray}
{\vec s}&=&\big(\sin\beta_s\cos\phi_s,\sin\beta_s\sin\phi_s,\cos\beta_s\big).
\label{eq:note-s}
\end{eqnarray}
As a result, using the light trajectory parametrization $\vec r=\vec r(\vec b,\tau)$ from (\ref{eq:b0}), we obtain the total accumulated gravitational phase shift $\varphi(\vec b)$ that takes the form  (see details in  \cite{Turyshev-Toth:2021-multipoles}):
{}
\begin{eqnarray}
\varphi(\vec b) =\frac{2k}{c^2}\int^{\tau}_{\tau_0}  U({\vec b},\tau') d\tau'=
kr_g\ln 4k^2rr_0-2kr_g\Big(\ln kb + \sum_{\ell=2}^\infty\frac{J_\ell}{\ell}\Big(\frac{R}{b}\Big)^\ell \sin^\ell\beta_s\cos[\ell(\phi_\xi-\phi_s)]\Big).
\label{eq:Psi-in}
\end{eqnarray}

The first two  terms in (\ref{eq:Psi-in}) represent the phase shift due to the monopole term of the gravitational field of the lens, while the last term is that due to contributions from the infinite set of the zonal harmonics, $J_\ell$.

We recognize that in the case when lensing on a large optically opaque body is considered, part of the incident radiation is being absorbed by the body (see discussion of the fully-absorbing boundary conditions, e.g., in \cite{Turyshev-Toth:2017,Turyshev-Toth:2021-multipoles}).
This results in the fact that the radial components of the impact parameter in the integral (\ref{eq:amp-A}) varies as $b\in [R,\infty[$ with $R$ being the characteristic size of the lensing object, while its angular coordinate varies as $\phi_\xi\in [0, 2\pi]$. Furthermore, given the fact that $b\geq R$, in the case of a typical astrophysical body (i.e., star,  compact galaxy, etc.), the magnitude of the logarithmic term in  (\ref{eq:Psi-in}) is much larger than that of the $J_\ell$-term.

These considerations allow us to evaluate the radial integral in (\ref{eq:amp-A}) by the method of stationary phase.  As a result,  in the case of an axisymmetric lens, the complex amplitude of the EM field (\ref{eq:amp-A}) takes the form  (see \cite{Turyshev-Toth:2021-multipoles,Turyshev-Toth:2021-all-regions}):
{}
\begin{eqnarray}
A(\vec x) &=&
\sqrt{2\pi kr_g}e^{i\sigma_0}e^{ik(r_0+r+r_g\ln 4k^2rr_0)}B_{\tt al}(\vec x),
  \label{eq:amp-A2}
\end{eqnarray}
where $\sigma_0=-kr_g\ln kr_g/e-{\textstyle\frac{\pi}{4}}$ is constant, discussed in \cite{Abramovitz-Stegun:1965,Turyshev-Toth:2019}, and $B_{\tt al}(\vec x)$ is the  remaining angular integral of  complex amplitude of the EM field for an axisymmetric lens \cite{Turyshev-Toth:2021-multipoles,Turyshev-Toth:2021-all-regions} (denoted with a subscript ``al''):
{}
\begin{eqnarray}
B_{\tt al}(\vec x) &=&
\frac{1}{2\pi}\int_0^{2\pi} d\phi_\xi \exp\Big[-ik\Big(\sqrt{\frac{2r_g}{\tilde r}}\rho\cos(\phi_\xi-\phi)+
2r_g\sum_{\ell=2}^\infty \frac{J_\ell}{\ell} \Big(\frac{R}{\sqrt{2r_g\tilde r}}\Big)^\ell\sin^\ell\beta_s\cos[\ell(\phi_\xi-\phi_s)]\Big)\Big].
  \label{eq:B2}
\end{eqnarray}
Eq.~(\ref{eq:B2}) is a new diffraction integral formula that was first reported in \cite{Turyshev-Toth:2021-multipoles}. It extends previous wave-theoretical descriptions of gravitational lensing phenomena to the case of a lens with an arbitrary axisymmetric structure.

This quantity allows one to compute the point-spread function (PSF) of an axisymmetric lens in the strong interference region behind the lens \cite{Turyshev-Toth:2021-multipoles,Turyshev-Toth:2021-imaging}:
{}
\begin{eqnarray}
{\rm PSF}({\vec x})&=&|B_{\tt al}({\vec x})|^2.
 \label{eq:psf}
\end{eqnarray}
The PSF, which represents the impulse-response of the gravitational lens, is the fundamental expression that is used in practical models of gravitational lenses, especially when imaging is concerned \cite{Turyshev-Toth:2021-imaging}.

\subsection{Eikonal phase shift for an axisymmetric body}
\label{sec:eik-phase-axsym-eik}

To capture the difference between various Newtonian potential models, we introduce the multipolar eikonal phase shift, $\xi_b(\vec b,\vec s)$. This shift is acquired by the EM wave as it interacts with the extended gravitational field of an axisymmetric body in its vicinity.  In the majority of practically important cases, the contribution of the monopole term within any model of the gravitational potential will result in the same structure of the first term in the phase of (\ref{eq:B2}). To quantify the difference between the models, we focus  on the contributions of multipoles to the total gravitational phase shift that is responsible for the second term in the phase of (\ref{eq:B2}).

Therefore, the quantity of interest---the eikonal gravitational phase shift, $2\xi_b(\vec b,\vec s)$---is obtained by dropping the monopole term from (\ref{eq:Psi-in}),  or by presenting (\ref{eq:Psi-in}) as $\varphi(\vec b) =kr_g\ln 4k^2rr_0-2kr_g\ln kb+2\xi_b(\vec b,\vec s)$, which yields the following result \cite{Turyshev-Toth:2021-multipoles}:
{}
\begin{eqnarray}
\xi_b(\vec b,\vec s)=
-kr_g\sum_{\ell=2}^\infty\frac{J_\ell}{\ell}\Big(\frac{R}{b}\Big)^\ell \sin^\ell\beta_s\cos[\ell(\phi_\xi-\phi_s)].
\label{eq:eik-ph-axi*}
\end{eqnarray}
 This expression and its impact on the optical properties of an axisymmetric lens was studied extensively in \cite{Turyshev-Toth:2021-multipoles,Turyshev-Toth:2021-caustics,Turyshev-Toth:2021-imaging,Turyshev-Toth:2021-quartic}.

Our objective is to generalize this expression to the case of arbitrary gravitational fields, including fields with no axial symmetry. Given the fact that contribution of the monopole term in the potential $U(\vec x)$ in (\ref{eq:amp-A}) to to the phase delay $\varphi(\vec b)$ from (\ref{eq:Psi-in}) will be identical for all cases, the difference will be due to the form of the potential used to capture the nonspherical part of the mass distribution.  For this purpose, we explore the use of the STF formalism.

\section{Lensing by bodies of arbitrary composition}
\label{sec:arb-STF-mom}

The solution given by (\ref{eq:DB-sol-rho})--(\ref{eq:amp-A}) allows us to consider  extended lenses, relying on physically motivated lens models where the gravitational potential is given in its most generic form. For practical applications, the potential $U(\vec r)$  is typically expanded in terms of spherical harmonics:
{}
\begin{eqnarray}
U(\vec r)&=&\frac{GM}{r}\Big(1+\sum_{\ell=2}^\infty\sum_{k=0}^{+\ell}\Big(\frac{R}{r}\Big)^\ell P_{\ell k}(\cos\theta)(C_{\ell k}\cos k\phi+S_{\ell k}\sin k\phi)\Big)+
{\cal O}(c^{-4}),
\label{eq:pot_w_0sh}
\end{eqnarray}
where $P_{\ell k}$ are the associated Legendre polynomials \cite{Abramovitz-Stegun:1965}, while $C_{\ell k}$ and $S_{\ell k}$ are  the normalized spherical harmonic coefficients that characterize nonspherical contributions to the gravitational field.

Although the form (\ref{eq:pot_w_0sh}) is effective for many applications in geodesy, it is not technically convenient when we study light propagation in a gravitational field. In part, this is due to the fact that integration of the gravitational potential in (\ref{eq:amp-A}) is not well defined for potentials of the form (\ref{eq:pot_w_0sh}). Thus, alternative representations of $U(\vec x)$ are needed. In \cite{Turyshev-Toth:2021-multipoles}, we considered the case of axisymmetric bodies (summarized in Sec.~\ref{sec:eik-phase-axsym}). We now consider a generic potential in the form of an expansion of $U(\vec x)$  using  STF tensors.

\subsection{Computing the gravitational phase shift using STF tensors}
\label{sec:eik-STF}

In \cite{Turyshev-Toth:2021-multipoles}, we observed that expanding the potential  (\ref{eq:w-PN}) in terms of STF tensors \cite{Thorne:1980,Blanchet-Damour:1986,Blanchet-Damour:1989,Kopeikin:1997,Mathis-LePoncinLafitte:2007}  offers a viable alternative to (\ref{eq:pot_w_0sh}).  As was discussed in \cite{Thorne:1980},  the  scalar gravitational potential (\ref{eq:w-PN}) may equivalently be given in the following form:
{}
\begin{eqnarray}
U(\vec r)&=&G\int \frac{\rho({\vec r'})d^3{\vec r}'}{|{\vec r}-{\vec r'}|}=GM
\sum_{\ell=0}^\infty \frac{(-1)^\ell}{\ell!} {\cal T}^{<a_1...a_\ell>}\frac{\partial^\ell}{\partial x^{<a_1...}\partial x^{a_\ell>}}\Big(\frac{1}{r}\Big) + {\cal O}(c^{-4}),
\label{eq:pot_stf}
\end{eqnarray}
where $r=|{\vec r}|$, $M$ is the post-Newtonian mass of the body, and ${\cal T}^{<a_1...a_\ell>}$ are the normalized STF multipole moments, defined as
{}
\begin{eqnarray}
M&=&\int d^3{\vec r'}\, \rho({\vec r'}),\qquad
{\cal T}^{<a_1...a_\ell>}=\frac{1}{M}\int d^3{\vec r'}\, \rho({\vec r'})\, x'{}^{<a_1...a_\ell>},
\label{eq:mom}
\end{eqnarray}
where $x^{<a_1...a_\ell>}=x^{<a_1}x^{a_2...}x^{a_\ell>}\equiv \hat x^L$, while the angle  brackets $<...>$ and $\hat{x}$ denote the STF operator (see Appendix~\ref{sec:stf-comb}, were we present the coordinate combinations needed to compute the lowest STF moments). Note that in the static post-Newtonian case we may use the multipole moments normalized to the mass of the body $M$. Also, without loss of generality, we can choose the origin of the coordinate system at the center-of-mass of the gravitating system, allowing us to eliminate the dipole moment ${\cal T}^a$ from the expansion (\ref{eq:pot_stf}).

Using the identity \cite{Soffel-Han:2019},
{}
\begin{eqnarray}
\frac{\partial^\ell}{\partial x^{<a_1...}\partial x^{a_\ell>}}\Big(\frac{1}{r}\Big)&=& (-1)^\ell(2\ell-1)!!\frac{\hat n_{<a_1...a_\ell>}}{r^{\ell+1}},
\label{eq:pot_stf-dir}
\end{eqnarray}
the potential (\ref{eq:pot_stf}) may be given in the following form:
{}
\begin{eqnarray}
U(\vec r)&=& GM\sum_{\ell\geq 0}\frac{(2\ell-1)!!}{\ell !}{\cal T}_L\frac{\hat n_L}{r^{\ell+1}}.
\label{eq:pot_w_0STF}
\end{eqnarray}
The first few terms of (\ref{eq:pot_w_0STF}) or, equivalently, (\ref{eq:pot_stf}), are given as
{}
\begin{eqnarray}
U(\vec r)&=&GM\Big\{\frac{1}{r}+ \frac{3{\cal T}_{<ab>}}{2r^5}x^ax^b +\frac{5{\cal T}_{<abc>}}{2r^7}x^ax^bx^c+\frac{35{\cal T}_{<abcd>}}{8r^9}x^ax^bx^cx^d+{\cal O}(r^{-6})\Big\}.
\label{eq:pot_w_0STF2}
\end{eqnarray}
This Cartesian multipole expansion of the Newtonian gravitational potential is equivalent to expansion in terms of spherical harmonics (\ref{eq:pot_w_0sh}) \cite{Thorne:1980,Blanchet-Damour:1986,Blanchet-Damour:1989,Kopeikin:1997,Mathis-LePoncinLafitte:2007}. In fact, this expression may be used to establish the correspondence between ${\cal T}^{<a_1...a_\ell>}$ and $C_{\ell k}$ and $S_{\ell k}$ from (\ref{eq:pot_w_0sh}) (see Appending~\ref{sec:stf-sph-harm} for details on how to establish such correspondence).

To compute the gravitational phase shift, we use $U({\vec r})$ from (\ref{eq:pot_stf}) in the expression for the phase shift $\varphi(\vec b)$ in (\ref{eq:Psi-in}):
{}
\begin{eqnarray}
\varphi(\vec b) &=&\frac{2k}{c^2}\int^{\tau}_{\tau_0}  U({\vec b},\tau') d\tau'= kr_g\sum_{\ell=0}^\infty
\frac{(-1)^\ell}{\ell!} {\cal T}^{<a_1...a_\ell>} \int^{\tau}_{\tau_0} \frac{\partial^\ell}{\partial x^{<a_1...}\partial x^{a_\ell>}}\Big(\frac{1}{r}\Big) d\tau'.
\label{eq:delta-D*-av0WKB+1*}
\end{eqnarray}

With the  definitions for $\vec r$ from (\ref{eq:b0}), we may generalize the expression for a gradient ${\vec \nabla}$ by relating it to the derivatives parallel to $\vec k$ and perpendicular to it, $\vec b$, namely ${\vec \nabla}={\nabla}^\perp+{\nabla}^\parallel \equiv {\nabla}_b+{\vec k}\,d/d\tau +{\cal O}(r_g)$ and write
{}
\begin{eqnarray}
\frac{\partial^\ell}{\partial x^{<a_1...}\partial x^{a_\ell>}}&\equiv&
{\vec \nabla}^{<a_1....}{\vec \nabla}^{a_\ell>}=\sum_{p=0}^\ell \frac{\ell!}{p!(\ell-p)!}k_{<a_1}...k_{a_p} \hat\partial_{a_{p+1}}... \hat\partial_{a_\ell>} \frac{\partial^p}{\partial \tau^p}+{\cal O}(r_g),
\label{eq:derivA}
\end{eqnarray}
where we use a new shorthand notation $\hat\partial_a\equiv \partial/\partial b^a$, with the hatted notation indicating differentiation that is carried out in two dimensions only, with respect to the two nonzero components of the impact parameter (see also Appendix \ref{sec:cases}), and $\tau$ is defined by (\ref{eq:x-Newt*=0}).

With this representation (\ref{eq:derivA}), we can compute the relevant integral (where from (\ref{eq:b0}) we have $r=\sqrt{b^2+\tau^2}$):
 {}
\begin{eqnarray}
 \int^{\tau}_{\tau_0} \frac{\partial^\ell}{\partial x^{<a_1...}\partial x^{a_\ell>}}\Big(\frac{1}{r}\Big) d\tau'&=&
 \sum_{p=0}^\ell \frac{\ell!}{p!(\ell-p)!}k_{<a_1}...k_{a_p} \hat\partial_{a_{p+1}}... \hat\partial_{a_\ell>} \int^{\tau}_{\tau_0}\frac{\partial^p}{\partial \tau'^p}\Big( \frac{1}{\sqrt{b^2+\tau'^2}}\Big)d\tau'.
\label{eq:int+1*}
\end{eqnarray}

Using (\ref{eq:x-Newt*=0}) and (\ref{eq:b0}), we evaluate the integral over $\tau'$ in (\ref{eq:int+1*}) by taking into account the fact that $\tau$ changes the sign after passing through $\tau=0$, being negative for $\tau_0$ to 0 and positive from  $0$ to $\tau$:
{}
\begin{eqnarray}
 \int_{\tau_0}^\tau\frac{d\tau'}{\sqrt{b^2+\tau'^2}} &=&
 \ln \Big(\frac{\sqrt{b^2+\tau^2}+\tau}{b}\Big)+
 \ln \Big(\frac{\sqrt{b^2+\tau_0^2}+|\tau_0|}{b}\Big),
\label{eq:int-ln}
\end{eqnarray}
where we employed the useful relations
{}
\begin{eqnarray}
\Big(\sqrt{b^2+\tau^2}+\tau\Big)\Big(\sqrt{b^2+\tau^2}-\tau\Big)=b^2\qquad
{\rm and}\qquad
\Big(\sqrt{b^2+\tau_0^2}+\tau_0\Big)\Big(\sqrt{b^2+\tau_0^2}-\tau_0\Big)=b^2.
\label{eq:b02}
\end{eqnarray}

As a result, (\ref{eq:int+1*}) takes the form:
 {}
\begin{eqnarray}
 \int^{\tau}_{\tau_0} \frac{\partial^\ell}{\partial x^{<a_1...}\partial x^{a_\ell>}}\Big(\frac{1}{r}\Big) d\tau'&=&
 \hat\partial_{<a_1}... \hat\partial_{a_\ell>}\Big\{ \ln \Big(\frac{\sqrt{b^2+\tau^2}+\tau}{b}\Big)+\ln \Big(\frac{\sqrt{b^2+\tau_0^2}+|\tau_0|}{b}\Big)\Big\}+\nonumber\\
 &+&
  \sum_{p=1}^\ell \frac{\ell!}{p!(\ell-p)!}k_{<a_1}...k_{a_p} \hat\partial_{a_{p+1}}... \hat\partial_{a_\ell>}
  \Big\{\frac{\partial^{p-1}}{\partial \tau^{p-1}}  \frac{1}{\sqrt{b^2+\tau^2}}+\frac{\partial^{p-1}}{\partial \tau_0^{p-1}}  \frac{1}{\sqrt{b^2+\tau_0^2}}\Big\},
\label{eq:int+3*3}
\end{eqnarray}
where we accounted for the fact that $\tau$ changes sign at $\tau=0$.

As a result, the gravitational phase shift (\ref{eq:delta-D*-av0WKB+1*}) takes the most general form:
{}
\begin{eqnarray}
\varphi(\vec b)
&=& kr_g\sum_{\ell=0}^\infty
\frac{(-1)^\ell}{\ell!} {\cal T}^{<a_1...a_\ell>}
\Big\{
 \hat\partial_{<a_1}... \hat\partial_{a_\ell>}\Big\{ \ln \Big(\frac{\sqrt{b^2+\tau^2}+\tau}{b}\Big)+\ln \Big(\frac{\sqrt{b^2+\tau_0^2}+|\tau_0|}{b}\Big)\Big\}+\nonumber\\
 &&\hskip 60pt\,+
  \sum_{p=1}^\ell \frac{\ell!}{p!(\ell-p)!}k_{<a_1}...k_{a_p} \hat\partial_{a_{p+1}}... \hat\partial_{a_\ell>}
  \Big\{\frac{\partial^{p-1}}{\partial \tau^{p-1}}  \frac{1}{\sqrt{b^2+\tau^2}}+\frac{\partial^{p-1}}{\partial \tau_0^{p-1}}  \frac{1}{\sqrt{b^2+\tau_0^2}}\Big\}\Big\}.
\label{eq:delta-eik-gen}
\end{eqnarray}

In the case of gravitational lensing, the typical distances involved are much larger than the impact parameter, $b/r\ll1$, and also $b/r_0\ll 1$, allowing us to simplify expression (\ref{eq:delta-eik-gen}). For that, we observe that $b/\sqrt{b^2+\tau^2}\equiv b/r\ll1$, $b/\sqrt{b^2+\tau_0^2}\equiv b/r_0\ll1$,  and also $(\vec k \cdot\vec r)\simeq r$, $(\vec k \cdot\vec r_0)\simeq -r_0$, thus the following approximations are valid:
{}
\begin{eqnarray}
\ln \Big(k(\sqrt{b^2+\tau^2}+\tau)\Big)&=&\ln [k(r+(\vec k \cdot\vec r))]\simeq \ln 2kr,
\nonumber\\
\quad
\ln \Big(k(\sqrt{b^2+\tau_0^2}+|\tau_0|)\Big)&=&\ln [k(r_0+|(\vec k \cdot\vec r_0)|)]\simeq \ln 2kr_0.
\label{eq:int-ln2}
\end{eqnarray}

We also observe that any derivatives in (\ref{eq:delta-eik-gen}) with respect to $b^{a_\ell}$ would result in expressions that are at most of ${\cal O}(b/r^2)$ or ${\cal O}(b/r^2_0)$, which are small and may be neglected.  We also realize that any $(p-1)$-th derivative, either with respect to $\tau$ or $\tau_0$, applied to the second term in (\ref{eq:delta-eik-gen}), would result in producing terms of ${\cal O}(1/r^p)$ or ${\cal O}(1/r_0^p)$, which are also small and may be neglected.  These considerations allow us to greatly simplify (\ref{eq:int+3*3}):
 {}
\begin{eqnarray}
 \int^{\tau}_{\tau_0} \frac{\partial^\ell}{\partial x^{<a_1...}\partial x^{a_\ell>}}\Big(\frac{1}{r}\Big) d\tau'&=&- 2\hat\partial_{<a_1}... \hat\partial_{a_\ell>}\ln kb,
\label{eq:int+37*}
\end{eqnarray}
and also the gravitational eikonal phase shift (\ref{eq:delta-eik-gen}) that may now be given as
 {}
\begin{eqnarray}
\varphi(\vec b)&=& kr_g\ln 4k^2rr_0 -2kr_g\Big(\ln kb+\sum_{\ell=2}^\infty
\frac{(-1)^\ell}{\ell!} {\cal T}^{<a_1...a_\ell>}
 \hat\partial_{<a_1}... \hat\partial_{a_\ell>}\ln kb \Big)+{\cal O}(r_g^2).
\label{eq:eik-ph-shift}
\end{eqnarray}

Similarly to the case of (\ref{eq:amp-A2})--(\ref{eq:B2}), we substitute result (\ref{eq:eik-ph-shift}) into (\ref{eq:amp-A}) and compute the integral with respect to the radial variable $b$, relying on the method of stationary phase while considering that the monopole term is dominant in the regions of interest (as we did in \cite{Turyshev-Toth:2021-multipoles}). As a result, in the case of an extended lens with arbitrary mass distribution, the complex amplitude of the EM field $B(\vec x) $ from (\ref{eq:amp-A2}) takes the form
{}
\begin{eqnarray}
B(\vec x) &=&
\frac{1}{2\pi}\int_0^{2\pi} d\phi_\xi \exp\Big[-ik\Big(\sqrt{\frac{2r_g}{r}}\rho\cos(\phi_\xi-\phi)+
2r_g\sum_{\ell=2}^\infty
\frac{(-1)^\ell}{\ell!} {\cal T}^{<a_1...a_\ell>} \big(
 \hat\partial_{<a_1}... \hat\partial_{a_\ell>}\ln kb\big)\Big|_{b=\sqrt{2r_gr}}\Big)\Big].~~~
  \label{eq:B2-ext}
\end{eqnarray}

We can use this result to derive the eikonal phase for any of the terms in the Newtonian potential (\ref{eq:pot_w_0STF}) and present in (\ref{eq:eik-ph-shift}) and (\ref{eq:B2-ext}).

Similarly to (\ref{eq:eik-ph-axi*}), we isolate the contribution of the STF multipole moments to the total phase shift $\varphi(\vec b)$ in (\ref{eq:eik-ph-shift}). This allows us to present the eikonal phase shift $\xi_b(\vec b)$, expressed via the STF multipole moments, as
 {}
\begin{eqnarray}
\xi_b(\vec b)&=& -kr_g\sum_{\ell=2}^\infty
\frac{(-1)^\ell}{\ell!} {\cal T}^{<a_1...a_\ell>}
 \hat\partial_{<a_1}... \hat\partial_{a_\ell>}\ln kb +{\cal O}(r_g^2),
\label{eq:eik-ph2}
\end{eqnarray}
which is the generalization of the eikonal phase shift (\ref{eq:eik-ph-axi*}). Note that, as opposed to (\ref{eq:eik-ph-axi*}), expression (\ref{eq:eik-ph2}) is given for a lensing object that is characterized by an arbitrary mass distribution. This is a key result. Together with (\ref{eq:B2-ext}), it allows us to describe gravitational lensing by the most general compact mass distribution.

\subsection{Gravitational phase shift for the lowest STF moments}
\label{sec:compute-shift}

To demonstrate the practical utility of our results, we now compute several low-order terms in (\ref{eq:eik-ph2}), for $\ell=2,3,4$.
In Appendix~\ref{sec:cases}, we compute the corresponding derivatives with respect to the vector impact parameter, which are present in (\ref{eq:eik-ph2}).  Below, we present the results for the eikonal phase shift for the quadrupole ($\ell=2$), octupole ($\ell=3$) and hexadecapole ($\ell=4$) STF multipole moments.

\subsubsection{Quadrupole moment}
\label{sec:quad-mom}

With the expression for  $\hat\partial^2_{ab}\ln kb$ from (\ref{eq:dab2}), we present the result for the eikonal phase $\xi^{[2]}_b(\vec b)$ characterizing the contribution from the quadrupole (i.e., $\ell=2$) STF mass moment, ${\cal T}^{<ab>}$:
{}
\begin{eqnarray}
\xi^{[2]}_b(\vec b)
&=& kr_g
\frac{1}{2b^2} {\cal T}^{<ab>}
\Big(2m^a m^b+k^ak^b-\delta^{ab}\Big),
\label{eq:eik-ph22*}
\end{eqnarray}
where $m^a=b^a/b$ is the $a$-th component of the unit vector in the direction of the impact parameter, see (\ref{eq:note-b}). Note that the expression in parentheses in (\ref{eq:eik-ph22*}), which came from the derivatives $ \hat\partial_{<ab>}\ln kb$, is already in STF form. It acts on the STF mass quadrupole moment ${\cal T}^{<ab>}$, projecting the quadrupole moment onto the plane perpendicular to $\vec k$.

We parameterize the vectors $\vec k$, $\vec m$ using (\ref{eq:note-k}), (\ref{eq:note-b}) and present  (\ref{eq:eik-ph22*}):
{}
\begin{eqnarray}
\xi^{[2]}_b(\vec b)&=& kr_g\frac{1}{2b^2}
\Big( ({\cal T}_{11}-{\cal T}_{22}) \cos2\phi_\xi
+2{\cal T}_{12} \sin2\phi_\xi\Big),
\label{eq:eik-ph22p*}
\end{eqnarray}
where the form of the quadrupole STF mass moment is kept in its generic form.

The rank-2 STF tensor ${\cal T}^{<ab>}$ in three dimensions has five independent components. Examining (\ref{eq:eik-ph22p*}), we see that when the distances $r_0$ and $r$ are large (i.e., the thin lens approximation), the eikonal phase shift is given only by its transverse part, which depends on the combination of three independent components of the quadrupole moment tensor.  The remaining two components, ${\cal T}_{13} $ and ${\cal T}_{23}$, are present only in the longitudinal part of the phase shift (i.e., in the direction parallel to $\vec k$, as opposed to the transverse components that are in the direction perpendicular to $\vec k$, see Sec.~\ref{sec:TT-proj}), hence they are not seen in lensing observations. This is due to the fact that in (\ref{eq:eik-ph22*}) the quadrupole STF mass moment ${\cal T}^{<ab>}$  is multiplied by $\big(2m^a m^b+k^ak^b-\delta^{ab}\big)$, which is a projection operator onto the plane of the impact parameter. Therefore, we observe only transverse components of ${\cal T}^{<ab>}$; effects due to longitudinal components are suppressed. As we demonstrate below, this behavior also characterizes higher-rank STF multipole tensors: in (\ref{eq:eik-ph2}) only the transverse parts of these tensors contribute to lensing observations for higher values of $\ell$.

In the case when distances are of the same order of magnitude, $b\sim r\sim r_0$, the result for $\varphi(\vec b)$ takes a more complicated form that depends on all the components of the ${\cal T}^{<ab>}$, both transverse and longitudinal. In Appendix~\ref{sec:beyond}, we derive the form of the eikonal phase shift for such cases, to show that although the expression formally depends on all the components of the STF quadrupole moment,  the longitudinal components are strongly suppressed by various powers of $b/r, b/r_0$, which in the case of astrophysical lenses are negligibly small, $b/r \ll 1, b/r_0\ll1$. This justifies the approximation that we use to derive (\ref{eq:eik-ph-shift}).

To further analyze (\ref{eq:eik-ph22p*}), we introduce two quantities:
{}
\begin{eqnarray}
Q_2&=&\sqrt{ ({\cal T}_{11}-{\cal T}_{22})^2+ 4{\cal T}^2_{12}},\qquad \cos2\phi_2=\frac{{\cal T}_{11}-{\cal T}_{22}}{Q_2},\qquad \sin2\phi_2=\frac{2{\cal T}_{12}}{Q_2},
\label{eq:eik-defQ2}
\end{eqnarray}
and present (\ref{eq:eik-ph22p*}) in a much simpler form:
{}
\begin{eqnarray}
\xi^{[2]}_b(\vec b)&=& kr_g\frac{Q_2}{2b^2}
 \cos[2(\phi_\xi-\phi_2)].
\label{eq:eik-phas2q}
\end{eqnarray}

We can see that the contribution of the quadrupole STF moment ${\cal T}^{<ab>}$  is reduced to just two parameters: the magnitude $Q_2$ and the phase $\phi_2$ that determine the size and the rotation angle of the quadrupole caustic formed in the PSF of a lens (see \cite{Turyshev-Toth:2021-multipoles,Turyshev-Toth:2021-caustics} for details).

There is a connection between ${\cal T}^{<ab>}$ in (\ref{eq:pot_w_0STF}) and the spherical harmonics present in (\ref{eq:pot_w_0sh}) (see the derivation given in Appendix~\ref{sec:stf-sph-harm}):
{}
\begin{eqnarray}
{\cal T}_{11}&=&\Big(-{\textstyle\frac{1}{3}}C_{20}+2C_{22}\Big)R^2, \qquad  {\cal T}_{12}=2S_{22}R^2, \nonumber\\
{\cal T}_{22}&=&\Big(-{\textstyle\frac{1}{3}}C_{20}-2C_{22}\Big)R^2, \qquad {\cal T}_{13}=-C_{21}R^2,\nonumber\\
{\cal T}_{33}&=&{\textstyle\frac{2}{3}} C_{20} R^2, \qquad\qquad \qquad\quad~~
{\cal T}_{23}=-S_{21}R^2.
\label{eq:sp-harm2}
\end{eqnarray}
Consistent with the vanishing trace of ${\cal T}_{ab}$, only five of these terms are independent. Using these quantities we compute (\ref{eq:eik-defQ2}):
{}
\begin{eqnarray}
Q_2&=&4R^2\sqrt{C_{22}^2+ S^2_{22}},\qquad \cos2\phi_2=\frac{C_{22}}{\sqrt{ C_{22}^2+ S^2_{22}}},\qquad \sin2\phi_2=\frac{S_{22}}{\sqrt{ C_{22}^2+ S^2_{22}}},
\label{eq:eik-defQ2*}
\end{eqnarray}
which can be used to express (\ref{eq:eik-phas2q}) as
{}
\begin{eqnarray}
\xi^{[2]}_b(\vec b)&=& kr_g\frac{2R^2}{b^2}\sqrt{C_{22}^2+ S^2_{22}}
 \cos[2(\phi_\xi-\phi_2)].
\label{eq:eik-phas2q*}
\end{eqnarray}

Considering the case of an axisymmetric body, we remember that  ${\cal T}^{<ab>}$ is given as
{}
\begin{eqnarray}
{\cal T}^{<ab>}&=&-MJ_2R^2\Big(s^as^b-\frac{1}{3}\delta^{ab}\Big),
\label{eq:sp-Tab}
\end{eqnarray}
where $\vec s$ is the vector of rotational axis given by (\ref{eq:note-s}). Using this expression, the result (\ref{eq:eik-ph22p*}) reduces to
{}
\begin{eqnarray}
\xi^{[2]}_b(\vec b)&=& -kr_gJ_2\frac{R^2}{2b^2}
 \sin^2\beta_s\cos[2(\phi_\xi-\phi_s)],
\label{eq:eik-ph22*-ss}
\end{eqnarray}
which is identical to the $\ell=2$ zonal harmonic term in the case of an axisymmetric mass distribution (\ref{eq:eik-ph-axi*}), which was studied in \cite{Turyshev-Toth:2021-multipoles,Turyshev-Toth:2021-caustics,Turyshev-Toth:2021-quartic}.  This correspondence shows the utility of the STF moments describing light propagation in the vicinity of a gravitating object with arbitrary mass distribution.

Comparing expressions (\ref{eq:eik-phas2q*}) and (\ref{eq:eik-ph22*-ss}), we see that they have nearly identical structure. However, (\ref{eq:eik-phas2q*}) was derived using the potential (\ref{eq:pot_w_0sh}), with spherical harmonic coefficients defined in a coordinate system with arbitrary orientation with respect to $\vec k$. As we see below, similar results can be obtained for $\ell>2$ due to the fact that  (\ref{eq:eik-phas2q*}) contains only the projection of the multipole moments on the plane transverse to  $\vec k$. By choosing the axis of rotation for each $\ell$, the remaining freedom is now fixed yielding (\ref{eq:eik-ph22*-ss}).

\subsubsection{Octupole moment}
\label{sec:oct-mom}

Setting $\ell=3$ in (\ref{eq:eik-ph2}),  we use the result for  $\hat\partial^3_{abc}\ln kb$ from (\ref{eq:dab3}) and derive the eikonal phase shift, $\xi^{[3]}_b(\vec b)$, introduced by the octupole STF moment, ${\cal T}^{<abc>}$, which is given as
{}
\begin{eqnarray}
\xi^{[3]}_b(\vec b)
&=& kr_g
\frac{1}{6b^3} {\cal T}^{<abc>}
\Big(8m^a m^b m^c -2m^a (\delta^{bc}-k^bk^c)-2m^b(\delta^{ac}-k^ak^c)- 2m^c(\delta^{ab}-k^ak^b)\Big).
\label{eq:eik-ph233*}
\end{eqnarray}
Once again we observe that the expression in parentheses in (\ref{eq:eik-ph233*}) that came from $ \hat\partial_{<abc>}\ln kb$ is already STF. It acts on the quantity ${\cal T}^{<abc>}$, also an STF tensor. As we shall see in Sec.~\ref{sec:TT-proj}, this results in a transverse traceless (TT) projection of ${\cal T}^{<abc>}$ onto the plane of the impact parameter that is perpendicular to $\vec k$.

Again using the parameterization for the vectors $\vec k$, $\vec m$ given in (\ref{eq:note-k}), (\ref{eq:note-b}), we present  (\ref{eq:eik-ph233*}) as
\begin{eqnarray}
\xi^{[3]}_b(\vec b)&=&
kr_g\frac{1}{6b^3} \Big( 2({\cal T}_{111}-3{\cal T}_{122}) \cos3\phi_\xi
+2(3{\cal T}_{112}-{\cal T}_{222}) \sin3\phi_\xi\Big).
\label{eq:eik-ph23*}
\end{eqnarray}

Similarly to the case of the quadrupole moment, we introduce two quantities:
{}
\begin{eqnarray}
Q_3&=&2\sqrt{ \big({\cal T}_{111}-3{\cal T}_{122}\big)^2+ \big(3{\cal T}_{112}-{\cal T}_{222}\big)^2},\quad
\cos3\phi_3=-\frac{2({\cal T}_{111}-3{\cal T}_{122})}{Q_3},\quad \sin3\phi_3=-\frac{2(3{\cal T}_{112}-{\cal T}_{222})}{Q_3},~~
\label{eq:eik-defQ3}
\end{eqnarray}
and  present (\ref{eq:eik-ph23*}) in a much simplified form:
{}
\begin{eqnarray}
\xi^{[3]}_b(\vec b)&=& -kr_g\frac{Q_3}{6b^3}
 \cos[3(\phi_\xi-\phi_3)].
\label{eq:eik-phas3q}
\end{eqnarray}

We observe that, yet again, the effect of the STF octupole moment ${\cal T}^{<abc>}$ is reduced to only two parameters: the scale $Q_3$ and the phase $\phi_3$, which determine the size and rotation angle of the caustic formed by the STF octupole contribution to the PSF.

In Appendix~\ref{sec:stf-sph-harm}, we present the correspondence between ${\cal T}^{<abc>}$ from (\ref{eq:pot_w_0STF}) and the spherical harmonics from (\ref{eq:pot_w_0sh}):
{}
\begin{eqnarray}
{\cal T}_{111}&=&\Big({\textstyle\frac{3}{5}}C_{31}-6C_{33}\Big)R^3,\qquad\quad
{\cal T}_{112}=\Big({\textstyle\frac{1}{5}}S_{31}-6S_{33}\Big)R^3,
\qquad\quad
{\cal T}_{113}=\Big(-{\textstyle\frac{1}{5}}C_{30}+2C_{32}\Big)R^3
\nonumber\\
{\cal T}_{122}&=&\Big({\textstyle\frac{1}{5}}C_{31}+6C_{33}\Big)R^3,\qquad\quad
{\cal T}_{222}=\Big({\textstyle\frac{3}{5}}S_{31}+6S_{33}\Big)R^3,
\qquad\quad
{\cal T}_{223}=\Big(-{\textstyle\frac{1}{5}}C_{30}-2C_{32}\Big)R^3
\nonumber\\
{\cal T}_{123}&=&2S_{32}R^3, \qquad\quad
{\cal T}_{133}=-{\textstyle\frac{4}{5}}C_{31}R^3,  \qquad\quad
{\cal T}_{233}=-{\textstyle\frac{4}{5}}S_{31}R^3,  \qquad\quad
{\cal T}_{333}={\textstyle\frac{2}{5}}C_{30}R^3.
\label{eq:sp-harm3}
\end{eqnarray}
It is easy to check that the rank-3 STF tensor ${\cal T}_{<abc>}$ has seven independent components; the values of the remaining 20 components are determined by its symmetries and vanishing trace.

Using these quantities we compute (\ref{eq:eik-defQ3})
{}
\begin{eqnarray}
Q_3&=&48R^3\sqrt{C_{33}^2+ S^2_{33}},\qquad \cos3\phi_3=\frac{C_{33}}{\sqrt{ C_{33}^2+ S^2_{33}}},\qquad \sin3\phi_3=\frac{S_{33}}{\sqrt{ C_{33}^2+ S^2_{33}}},
\label{eq:eik-defQ3*}
\end{eqnarray}
that  transforms (\ref{eq:eik-phas3q})  as
{}
\begin{eqnarray}
\xi^{[3]}_b(\vec b)&=& -kr_g\frac{8R^3}{b^3}\sqrt{C_{33}^2+ S^2_{33}}
 \cos[3(\phi_\xi-\phi_3)].
\label{eq:eik-phas3q*}
\end{eqnarray}

Also, we know that in the axisymmetric case, ${\cal T}^{<abc>}$ has the form
{}
\begin{eqnarray}
{\cal T}^{<abc>}&=&-MJ_3R^3\Big\{s^as^bs^c-\frac{1}{5}\delta^{ab}s^c-\frac{1}{5}\delta^{ac}s^b-\frac{1}{5}\delta^{bc}s^a\Big\}.
\label{eq:sp-Tabc}
\end{eqnarray}
With this, using the parametrization (\ref{eq:note-s}), expression (\ref{eq:eik-ph23*}) reduces to
{}
\begin{eqnarray}
\xi^{[3]}_b(\vec b)&=& -kr_gJ_3\frac{R^3}{3b^3}
\sin^3\beta_s\cos[3(\phi_\xi-\phi_s)],
\label{eq:eik-ph22*-sss}
\end{eqnarray}
which is identical to the term with $\ell=3$ in (\ref{eq:eik-ph-axi*}), studied in \cite{Turyshev-Toth:2021-multipoles,Turyshev-Toth:2021-caustics}. Similarly to the quadrupole case, expressions (\ref{eq:eik-phas3q*}) and (\ref{eq:eik-ph22*-sss}) are formally identical.

\subsubsection{Hexadecapole moment}
\label{sec:hexa-mom}

In the case when $\ell=4$, we use the result for  $\hat\partial^4_{abcd}\ln kb$ from (\ref{eq:dab4}) and derive the eikonal phase shift, $\xi^{[4]}_b(\vec b)$, introduced by the hexadecapole STF moment, ${\cal T}^{<abcd>}$, which has the form
{}
\begin{eqnarray}
\xi^{[4]}_b(\vec b)&=& kr_g
\frac{1}{4b^4} {\cal T}^{<abcd>}
\Big\{8m^a m^b m^c m^d+
\frac{1}{3}\Big(\delta^{bc}-k^bk^c\Big)\Big(\delta^{ad}-k^ak^d\Big)+
\frac{1}{3}\Big(\delta^{ac}-k^ak^c\Big)\Big(\delta^{bd}-k^bk^d\Big)+
\nonumber\\
&&\hskip 10pt +\,
\frac{1}{3}\Big(\delta^{ab}-k^ak^b\Big)\Big(\delta^{cd}-k^ck^d\Big)-
\frac{4}{3}\Big(m^am^b\big(\delta^{cd}-k^ck^d\big)+
m^am^c\big(\delta^{bd}-k^bk^d\big)+
m^am^d\big(\delta^{bc}-k^bk^c\big)+
\nonumber\\
&&\hskip 10pt +\,
m^bm^c\big(\delta^{ad}-k^ak^d\big)+m^b m^d\big(\delta^{ac}-k^ak^c\big)+m^c m^d\big(\delta^{ab}-k^ak^b\big)\Big)\Big\}.
\label{eq:eik-ph244*}
\end{eqnarray}
Similarly to the cases of quadrupole and octopole considered earlier, the expression in parentheses in (\ref{eq:eik-ph244*}) that came from $ \hat\partial_{<abcd>}\ln kb$ is in an STF form. It  acts on the quantity ${\cal T}^{<abcd>}$, that is an STF tensor. This again results in the TT-projection of ${\cal T}^{<abcd>}$ onto the plane of the impact parameter that is perpendicular to $\vec k$.

Again, using  parameterizations for vectors $\vec k$, $\vec m$, from (\ref{eq:note-k}), (\ref{eq:note-b}), we present  (\ref{eq:eik-ph244*}) as
{}
\begin{eqnarray}
\xi^{[4]}_b(\vec b)&=&
kr_g
\frac{1}{4b^4} \Big\{ ({\cal T}_{1111}+{\cal T}_{2222}-6{\cal T}_{1122}) \cos4\phi_\xi
+4({\cal T}_{1112}-{\cal T}_{1222}) \sin4\phi_\xi\Big\}.
\label{eq:eik-ph24*}
\end{eqnarray}

Introducing the two quantities
{}
\begin{eqnarray}
Q_4&=&6\sqrt{ \big({\cal T}_{1111}+{\cal T}_{2222}-6{\cal T}_{1122} \big)^2+ 16\big({\cal T}_{1112}-{\cal T}_{1222}\big)^2},\\ \cos4\phi_4&=&\frac{6({\cal T}_{1111}+{\cal T}_{2222}-6{\cal T}_{1122})}{Q_4},\qquad \sin4\phi_4=\frac{6(4({\cal T}_{1112}-{\cal T}_{1222}))}{Q_4},
\label{eq:eik-defQ4}
\end{eqnarray}
we present (\ref{eq:eik-ph24*}) in a much simplified form:
{}
\begin{eqnarray}
\xi^{[4]}_b(\vec b)&=& kr_g\frac{Q_4}{24b^4}
 \cos[4(\phi_\xi-\phi_4)].
\label{eq:eik-phas4q}
\end{eqnarray}

Once again, the hexadecapole moment ${\cal T}^{<abcd>}$ is reduced to only two parameters, $Q_4$ and $\phi_4$, representing the two degrees of freedom available in the impact parameter plane.

Following the method outlined in Appendix~\ref{sec:stf-sph-harm}, we can establish a correspondence between ${\cal T}^{<abcd>}$ from (\ref{eq:pot_w_0STF}) and the spherical harmonics present in (\ref{eq:pot_w_0sh}):
{}
\begin{eqnarray}
{\cal T}_{1111}&=&\Big({\textstyle\frac{3}{35}}C_{40}- {\textstyle\frac{12}{7}}C_{42}+24C_{44}\Big)R^4,\qquad\quad
{\cal T}_{2222}=\Big({\textstyle\frac{3}{35}}C_{40}+{\textstyle\frac{12}{7}}C_{42}+24C_{44}\Big)R^4,
\nonumber\\
{\cal T}_{1112}&=&\Big(-{\textstyle\frac{6}{7}}S_{42}+ 24S_{44}\Big)R^4,\qquad\quad~\,
{\cal T}_{1113}=\Big({\textstyle\frac{3}{7}}C_{41}-{\textstyle\frac{60}{7}}C_{43}\Big)R^4,
\qquad\quad
{\cal T}_{2223}=\Big({\textstyle\frac{3}{7}}S_{41}+{\textstyle\frac{60}{7}}S_{43}\Big)R^4,
\nonumber\\
{\cal T}_{1122}&=&\Big({\textstyle\frac{1}{35}}C_{40}- 24C_{44}\Big)R^4,
\qquad\quad~~~\,
{\cal T}_{1123}=\Big({\textstyle\frac{1}{7}}S_{41}-{\textstyle\frac{60}{7}}S_{43}\Big)R^4,
\qquad\quad\,\,
{\cal T}_{1133}=\Big(-{\textstyle\frac{4}{35}}C_{40}+{\textstyle\frac{12}{7}}C_{42}\Big)R^4,
\nonumber\\
{\cal T}_{2233}&=&\Big(-{\textstyle\frac{4}{35}}C_{40}-{\textstyle\frac{12}{7}}C_{42}\Big)R^4, \qquad\quad
{\cal T}_{1222}=\Big(-{\textstyle\frac{6}{7}}S_{42}- 24S_{44}\Big)R^4,
\qquad\,
{\cal T}_{1223}=\Big({\textstyle\frac{1}{7}}C_{41}+{\textstyle\frac{60}{7}}C_{43}\Big)R^4,
\nonumber\\
{\cal T}_{3333}&=&{\textstyle\frac{8}{35}}C_{40}R^4, \qquad\quad
{\cal T}_{1233}={\textstyle\frac{12}{7}}S_{42}R^4,  \qquad\quad
{\cal T}_{1333}=-{\textstyle\frac{4}{7}}C_{41}R^4,  \qquad\quad
{\cal T}_{2333}=-{\textstyle\frac{4}{7}}S_{41}R^4.
\label{eq:sp-harm4}
\end{eqnarray}
As these quantities are components of an STF tensor, out of the fifteen terms ${\cal T}^{<abcd>}$ in (\ref{eq:sp-harm4}), only nine are independent.

Using these quantities we compute (\ref{eq:eik-defQ4}):
{}
\begin{eqnarray}
Q_4&=&1152R^4\sqrt{C_{44}^2+ S^2_{44}},\qquad \cos4\phi_4=\frac{C_{44}}{\sqrt{ C_{44}^2+ S^2_{44}}},\qquad \sin4\phi_4=\frac{S_{44}}{\sqrt{ C_{44}^2+ S^2_{44}}},
\label{eq:eik-defQ4*}
\end{eqnarray}
and transform (\ref{eq:eik-phas4q})  as
{}
\begin{eqnarray}
\xi^{[4]}_b(\vec b)&=& kr_g\frac{48R^4}{b^4}\sqrt{C_{44}^2+ S^2_{44}}
 \cos[4(\phi_\xi-\phi_4)].
\label{eq:eik-phas4q*}
\end{eqnarray}

In the axisymmetric case, we can calculate the STF moment ${\cal T}^{abcd}$ as
{}
\begin{eqnarray}
{\cal T}^{<abcd>}&=&-MJ_4R^4\Big\{s^as^bs^cs^d-\frac{1}{7}\Big(s^as^b\delta^{cd}+s^as^c\delta^{bd}+s^as^d\delta^{bc}+s^bs^c\delta^{ad}+s^bs^d\delta^{ac}+s^cs^d\delta^{ab}\Big)+
\nonumber\\
&& \hskip 85pt +\,\frac{1}{35}\Big(\delta^{ab}\delta^{cd}+\delta^{ac}\delta^{bd}+\delta^{ad}\delta^{bc}\Big)\Big\}.
\label{eq:sp-Tabcb}
\end{eqnarray}
With this form of ${\cal T}^{<abcd>}$, using the parametrization (\ref{eq:note-s}), expression (\ref{eq:eik-ph24*}) reduces to:
{}
\begin{eqnarray}
\xi^{[4]}_b(\vec b)&=& -kr_gJ_4\frac{R^4}{4b^4}
\sin^4\beta_s\cos[4(\phi_\xi-\phi_s)],
\label{eq:eik-ph22*-ssss}
\end{eqnarray}
which is identical to the term with $\ell=4$ in expression (\ref{eq:eik-ph-axi*}) that was discussed in \cite{Turyshev-Toth:2021-multipoles,Turyshev-Toth:2021-caustics}.

Once again, we see that the two forms of the gravitational phase shift (\ref{eq:eik-phas4q*}) and (\ref{eq:eik-ph22*-ssss}) are formally identical.

\section{Lensing with a generic gravitational lens}
\label{sec:generic}

As we saw in Sec.~\ref{sec:compute-shift}, even at the $\ell=4$ representing the hexadecapole moment, the gravitational phase shift is represented by a familiar expression that is determined by only two parameters: a magnitude and a rotation angle. This is because the quantity $\hat\partial_{<a_1}... \hat\partial_{a_\ell>}\ln kb$ in (\ref{eq:eik-ph2}) behaves as the transverse projection operator, thus projecting the STF moment of the appropriate order onto the plane of the impact parameter, i.e., the lens plane.  Therefore, the product of ${\cal T}^{<a_1...a_\ell>}  \hat\partial_{<a_1}... \hat\partial_{a_\ell>}\ln kb$,  in (\ref{eq:eik-ph2}), may be generalized and given as
{}
\begin{eqnarray}
 {\cal T}^{<a_1...a_\ell>} \hat\partial_{<a_1}... \hat\partial_{a_\ell>}\ln kb=(-1)^{\ell+1}\frac{Q_\ell}{b^\ell}\cos[\ell(\phi_\xi-\phi_\ell)],
  \label{eq:def-angl*-ex=}
\end{eqnarray}
where $Q_\ell$ and $\phi_\ell$ are the magnitude and the rotation angle that are computed or estimated for each order $\ell$ of the STF multiple moments from their subset projected on the plane of the impact parameter.

\subsection{Generalized expression for the gravitational phase shift}
\label{sec:TT-proj}

In Appendix~\ref{sec:projops}, we compute the derivatives of the $\hat\partial_{<a_1...a_\ell>}\ln kb$, (\ref{eq:tt23*f})--(\ref{eq:tt43*f}). Each of derivative $\ln kb$ with respect to the vector impact parameter produces an appropriate TT projection operator that is multiplied by various powers of $m^a$ and Kronecker delta. (An interesting observation is that these expressions are STF in two-dimensions, but not STF in three dimensions.) In fact,  the following expressions are valid:
{}
\begin{eqnarray}
 {\cal T}^{<ab>} \hat\partial_{<ab>}\ln kb&=& -\big[{\cal T}^{<ij>}\big]^{\rm TT}\frac{2}{b^2}\,m^im^j,
  \label{eq:tt23}\\
  {\cal T}^{<abc>} \hat\partial_{<abc>}\ln kb&=&
  \big[{\cal T}^{<ijk>}\big]^{\rm TT}\frac{8}{b^3}\,m^im^jm^k,
  \label{eq:tt33} \\
   {\cal T}^{<abcd>} \hat\partial_{<abcd>}\ln kb&=&-
\big[{\cal T}^{<ijkl>}\big]^{\rm TT}\frac{48}{b^4}\,m^im^jm^km^l,
  \label{eq:tt43}
\end{eqnarray}
where the quantities $\big[{\cal T}^{<a_1...a_\ell>}\big]^{\rm TT}$ are the transverse traceless (TT) projections of the STF multipole moments onto the plane of the impact parameter, perpendicular to $\vec k$.

Considering the structure of the derivatives $\hat\partial_{<a_1...a_\ell>}\ln kb$, we can see that this behavior persists at each order $\ell$, yielding a corresponding TT projection operator. Therefore, based on these considerations,  expressions $ {\cal T}^{<a_1...a_\ell>} \hat\partial_{<a_1...a_\ell>}\ln kb$ may be generalized to arbitrary order:
{}
\begin{eqnarray}
 {\cal T}^{<a_1...a_\ell>} \hat\partial_{<a_1...a_\ell>}\ln kb&=& (-1)^{\ell+1}\frac{(2\ell-2)!!}{b^\ell}\big[{\cal T}^{<a_1....a_\ell>}\big]^{\rm TT}\,m^{a_1...}m^{a_\ell}.
  \label{eq:tt23*}
\end{eqnarray}

Next, examining the structure of the TT-projected STF multipole moments, $\big[{\cal T}^{<a_1...a_\ell>}\big]^{\rm TT}$, we see that at each order they are given by distinct combinations:
{}
\begin{eqnarray}
\big[{\cal T}^{<a_1....a_\ell>}\big]^{\rm TT}=t^+_\ell U^+_{a_1...a_\ell} +t^\times_\ell U^\times_{a_1...a_\ell},
  \label{eq:tt5*a}
\end{eqnarray}
where $t^+_\ell$ and $t^\times_\ell$ are the components of the  $\big[{\cal T}^{<a_1....a_\ell>}\big]^{\rm TT}$ tensor and $U^+_{a_1...a_\ell} $ and $U^\times_{a_1...a_\ell} $ are the two orthogonal basis vectors for each order (similar to the two polarizations that exist in the quadrupole formalism of generating gravitational waves \cite{Thorne:1980}, which  usually is done only at $\ell=2$ level). The lowest order $t^+_\ell$ and $t^\times_\ell$ are given as
{}
\begin{eqnarray}
t^+_2&=&{\textstyle\frac{1}{2}} ({\cal T}_{11}-{\cal T}_{22}), \qquad\qquad\qquad\quad~~
t^\times_2= {\cal T}_{12},
\label{eq:eik-tt2}\\
t^+_3&=&{\textstyle\frac{1}{4}}({\cal T}_{111}-3{\cal T}_{122}),
\qquad\qquad\qquad~
 t^\times_3={\textstyle\frac{1}{4}} (3{\cal T}_{112}-{\cal T}_{222}),
\label{eq:eik-tt3}\\
t^+_4&=&{\textstyle\frac{1}{8}}({\cal T}_{1111}+{\cal T}_{2222}-6{\cal T}_{1122}) , \qquad   t^\times_4={\textstyle\frac{1}{2}} ({\cal T}_{1112}-{\cal T}_{1222}).
\label{eq:eik-tt4}
\end{eqnarray}

Concerning the STF basis vectors $U^+_{a_1...a_\ell} $ and $U^\times_{a_1...a_\ell} $, they represent the $\ell$-times rotation operators in the 2-dimensional plane perpendicular to $\vec k$. For convenience, explicit structure of $U^+_{a_1...a_\ell} $ and $U^\times_{a_1...a_\ell} $ for $\ell\in\{2,3,4\}$, is given in Appendix~\ref{sec:pol-mat}. One can verify that for the  parameterization (\ref{eq:note-k})--(\ref{eq:note-b}),  the following important relations exist:
{}
\begin{eqnarray}
U^+_{a_1...a_\ell}\,m^{a_1...}m^{a_\ell}=\cos[\ell\phi_\xi],
\qquad
U^\times_{a_1...a_\ell} \,m^{a_1...}m^{a_\ell}=\sin[\ell\phi_\xi].
  \label{eq:tt52q}
\end{eqnarray}
Thus, the basis vectors are the rotation generators that, in combination with the unit vectors $m^a$, act as the Chebychev polynomials \cite{Abramovitz-Stegun:1965}, rotating the products (\ref{eq:tt52q}) by the angle $\ell\phi_\xi$ in two orthogonal directions. Note that a similar result was obtained in \cite{Clarkson-2016-II} by integrating the non-linear geodesic deviation equation and generalizing the result to a sum over independent amplification tensors of increasing rank, thus validating our approach.

With the important property (\ref{eq:tt52q}), from (\ref{eq:tt5*a}) we derive another useful expression:
{}
\begin{eqnarray}
\big[{\cal T}^{<a_1....a_\ell>}\big]^{\rm TT}\,m^{a_1...}m^{a_\ell}&=&
t^+_\ell \cos[\ell\phi_\xi]+t^\times_\ell \sin[\ell\phi_\xi]=
\sqrt{t^{+2}_\ell +t^{\times2}_\ell}\cos[\ell(\phi_\xi-\phi_\ell)],
  \label{eq:tt5a7}
\end{eqnarray}
where the angle $\phi_\ell$ is determined from
{}
\begin{eqnarray}
\cos[\ell\phi_\ell]=\frac{t^+_\ell}{\sqrt{t^{+2}_\ell +t^{\times2}_\ell}},\qquad
\sin[\ell\phi_\ell]=\frac{t^\times_\ell}{\sqrt{t^{+2}_\ell +t^{\times2}_\ell}}.
  \label{eq:tt52a7}
\end{eqnarray}

As a result, the gravitational eikonal phase shift (\ref{eq:eik-ph2}) takes the following  compact form:
 {}
\begin{eqnarray}
\xi_b(\vec b)&=& kr_g\sum_{\ell=2}^\infty
\frac{(2\ell-2)!!}{\ell! \,b^\ell} \sqrt{t^{+2}_\ell +t^{\times2}_\ell}\cos[\ell(\phi_\xi-\phi_\ell)] +{\cal O}(r_g^2).
\label{eq:eik-ph2-STF}
\end{eqnarray}

We observe that at each order, the gravitational phase shift is determined by only the two degrees of freedom of the corresponding TT-projected STF multipole moment, $t^{+}_\ell $ and $t^{\times}_\ell$. The structure of the result (\ref{eq:eik-ph2-STF}) is very familiar to us from \cite{Turyshev-Toth:2021-multipoles}, where we studied the case of lenses with an axisymmetric  matter distribution.  What is surprising is the fact that even an arbitrary lens exhibits the same structure seen in (\ref{eq:eik-ph-axi*}). The difference is that at each STF order, $\ell$, the amplitude and the angle of (\ref{eq:tt52a7})--(\ref{eq:eik-ph2-STF}) is set by only two TT-projected STF mass multipole moments, $t^{+}_\ell$ and $t^{\times}_\ell$.

\subsection{Relation to spherical harmonics}

Using the correspondence between the STF mass moments at various orders and the spherical harmonic coefficients, given by (\ref{eq:sp-harm2}), (\ref{eq:sp-harm3}), and (\ref{eq:sp-harm4}), correspondingly, we have the following relationships:
{}
\begin{eqnarray}
t^+_2&=&2C_{22}R^2, \qquad\qquad~\,\,
t^\times_2= 2S_{22}R^2,
\label{eq:eik-tt12}\\
t^+_3&=&-6C_{33}R^3,
\qquad\qquad
 t^\times_3=-6S_{33}R^3,
\label{eq:eik-tt13}\\
t^+_4&=&24C_{44}R^4, \qquad\qquad\,   t^\times_4=24S_{44}R^4.
\label{eq:eik-tt14}
\end{eqnarray}

Again, one may generalize these expressions to arbitrary order, yielding
{}
\begin{eqnarray}
t^+_\ell&=&(-1)^\ell\ell! \,C_{\ell\ell}R^\ell, \qquad\qquad~\,\,
t^\times_2=(-1)^\ell\ \ell! \,S_{\ell\ell}R^\ell.
\label{eq:eik-tt12+}
\end{eqnarray}

As a result, (\ref{eq:tt5a7})--(\ref{eq:tt52a7}) transforms as
{}
\begin{eqnarray}
\big[{\cal T}^{<a_1....a_\ell>}\big]^{\rm TT}\,m^{a_1...}m^{a_\ell}
=\ell! R^\ell \sqrt{C^2_{\ell\ell} +S^2_{\ell\ell}}\cos[\ell(\phi_\xi-\phi_\ell)],
  \label{eq:tt5*}
\end{eqnarray}
where
{}
\begin{eqnarray}
\cos[\ell\phi_\ell]=\frac{(-1)^\ell C_{\ell\ell}}{\sqrt{C_{\ell\ell}^2+S^2_{\ell\ell}}},\qquad
\sin[\ell\phi_\ell]=\frac{(-1)^\ell S_{\ell\ell}}{\sqrt{C_{\ell\ell}^2+S_{\ell\ell}^2}}.
  \label{eq:tt52}
\end{eqnarray}

Ultimately, we express (\ref{eq:tt23*}) as
{}
\begin{eqnarray}
 {\cal T}^{<a_1...a_\ell>} \hat\partial_{<a_1...a_\ell>}\ln kb&=& (-1)^{\ell+1} \ell! (2\ell-2)!!
\Big(\frac{R}{b}\Big)^\ell \sqrt{C^2_{\ell\ell} +S^2_{\ell\ell}}\cos[\ell(\phi_\xi-\phi_\ell)].
  \label{eq:tt23*5}
\end{eqnarray}
With this result, the gravitational phase shift (\ref{eq:eik-ph2}) takes the form
 {}
\begin{eqnarray}
\xi_b(\vec b)&=& kr_g\sum_{\ell=2}^\infty
 (2\ell-2)!!
\Big(\frac{R}{b}\Big)^\ell \sqrt{C^2_{\ell\ell} +S^2_{\ell\ell}}\cos[\ell(\phi_\xi-\phi_\ell)] +{\cal O}(r_g^2).
\label{eq:eik-ph2-gen}
\end{eqnarray}

This is an important conclusion as it significantly simplifies the modeling of generic compact astrophysical lenses, including stars as well as distant and compact spiral and elliptic galaxies. The parameters $C_{\ell\ell}$ and $S_{\ell\ell}$ are very natural, and may be used to study arbitrary matter distributions. If these parameters are known, the task of developing physically justified  models for the corresponding lenses may be significantly simplified.

\subsection{Generalized expression for light deflection}

Using (\ref{eq:eik-ph-shift}), we identify  the total gravitational phase shift, $\varphi(\vec b)$, acquired by the EM wave as it travels through the gravitational field of the extended lens. This is a generalization of the classic Shapiro time delay to the case of an extended gravitational lens with arbitrary mass distribution. This delay corresponds to the total gravitational deflection angle acquired by a light ray or, equivalently, rotation of the wavefront of the EM wave. Using the expression (\ref{eq:b0}) for the radius vector of the EM wave, together with  $\vec b$ given by  (\ref{eq:note-s}), we compute this angle as
{}
\begin{eqnarray}
\vec \theta_g =-k^{-1}\vec \nabla \varphi(\vec b)=-k^{-1}\Big\{{\vec e}_b \frac{\partial  \varphi(\vec b)}{\partial b}+{\vec e}_{\phi_\xi} \frac{\partial  \varphi(\vec b)}{b\, \partial \phi_\xi}+{\vec k} \frac{\partial  \varphi(\vec b)}{\partial \tau}\Big\},
  \label{eq:grad-b*}
\end{eqnarray}
where the basis vector ${\vec e}_b$ is the unit vector in the direction of the vector impact parameter  $\vec{b}$ and ${\vec e}_{\phi_\xi}$ is the unit vector in the azimuthal direction and is orthogonal to ${\vec b}$ and ${\vec k}$. Clearly, in the case of a compact lens (i.e., neglecting the terms $b/r$ and $b/r_0$ in (\ref{eq:delta-eik-gen})) delay, computed from (\ref{eq:eik-ph-shift}), does not depend on $\tau$, thus the last derivative in (\ref{eq:grad-b*}) results in 0.

With these considerations in mind, we compute the vector of the total  angle of the gravitational deflection of light as the light ray passes in the vicinity of an extended gravitational lens with arbitrary mass distribution:
{}
\begin{eqnarray}
\vec \theta_g &=&2r_g\Big\{\frac{{\vec e}_b}{b}+\sum_{\ell=2}^\infty
\frac{(-1)^\ell}{\ell!} {\cal T}^{<a_1...a_\ell>} \Big({\vec e}_b \frac{\partial  }{\partial b}+{\vec e}_{\phi_\xi} \frac{\partial  }{b\, \partial \phi_\xi}\Big) \hat\partial_{<a_1}... \hat\partial_{a_\ell>}\ln kb\Big\}.
  \label{eq:def-angl*}
\end{eqnarray}

The first term in (\ref{eq:def-angl*}) is the Einstein deflection angle in the gravity field of a spherically symmetric matter distribution (i.e., in the presence of a monopole or point mass).  The second term with ${\cal T}^{<a_1...a_\ell>} $ describes the effect of the multipole STF moments as a sum of
\begin{inparaenum}[i)]
\item an additional deflection toward or away from the optical axis (the line parallel to the incoming ray of light that intersects the lens at the center), and
\item a deflection away from the plane defined by the incoming ray and the center of the lens.
\end{inparaenum}

As we discussed in this document, the product of ${\cal T}^{<a_1...a_\ell>}  \hat\partial_{<a_1}... \hat\partial_{a_\ell>}\ln kb$, present in (\ref{eq:def-angl*}), may be given by (\ref{eq:def-angl*-ex=}) with $Q_\ell$ and $\phi_\ell$ being the magnitude and the rotation angle computed for each order $\ell$ of the STF multipole moment, projected onto the plane of the impact parameter.  Using the result (\ref{eq:def-angl*-ex=}) in (\ref{eq:def-angl*}), we have
{}
\begin{eqnarray}
\vec \theta_g &=&\frac{2r_g}{b}\Big\{{\vec e}_b+\sum_{\ell=2}^\infty
\frac{(2\ell-2)!!}{(\ell-1)! \,b^\ell} \sqrt{t^{+2}_\ell +t^{\times2}_\ell}\Big({\vec e}_b \cos[\ell(\phi_\xi-\phi_\ell)]+{\vec e}_{\phi_\xi} \sin[\ell(\phi_\xi-\phi_\ell)]\Big) \Big\}.
  \label{eq:def-angl*4}
\end{eqnarray}

We note that the axisymmetric case with delay computed with the help of the corresponding phase shift (\ref{eq:Psi-in}) was considered in \cite{Turyshev-Toth:2021-multipoles}. The result given by (\ref{eq:def-angl*4}) is new.  It describes the three-dimensional deflection of light in the presence of an arbitrary gravitational field of an extended lens.

We recall that (\ref{eq:eik-ph-shift}) was obtained from  (\ref{eq:delta-eik-gen}) under the conditions $b/r_0\ll1$ and $b/r\ll1$. As a result, we are not sensitive to the longitudinal components of the gravitational field expressed via the STF multipole moments. In the case when $b/r_0\sim1$ and $b/r\sim1$, the phase shift (\ref{eq:delta-eik-gen}) contains all the components of the STF moments. In Appendix~\ref{sec:beyond}, we consider this for the quadrupole case, $\ell=2$. However, lensing geometries where the lens-centric distances to the emitter and observer are comparable to the size of the lens itself are rare. These mostly concern light propagation to and from interplanetary spacecraft within the solar system. For these few cases, the expressions given in  Appendix~\ref{sec:beyond} will be sufficient. However in the case of a generic lensing situation, expression (\ref{eq:def-angl*4}) represents the most comprehensive treatment for the gravitational deflection of light by extended gravitating bodies.

\subsection{Optical properties of a generic lens}

Using the gravitational eikonal phase shift given (\ref{eq:eik-ph-shift}), we can express (\ref{eq:eik-ph2-STF}) as
 {}
\begin{eqnarray}
\varphi(\vec b)&=& kr_g\ln 4k^2rr_0 -2kr_g\Big(\ln kb-\sum_{\ell=2}^\infty
\frac{(2\ell-2)!!}{\ell! \,b^\ell} \sqrt{t^{+2}_\ell +t^{\times2}_\ell}\cos[\ell(\phi_\xi-\phi_\ell)]  \Big)+{\cal O}(r_g^2).
\label{eq:eik-ph-shift_tpm}
\end{eqnarray}
Substituting this expression in (\ref{eq:amp-A}), we get
{}
\begin{eqnarray}
A(\vec x) &=&e^{ikr_g\ln 4k^2rr_0}
 \frac{k}{ir}\frac{1}{2\pi}\iint d^2\vec b \,\exp\Big[ik\Big(\frac{1}{2 \tilde r}({\vec b} - \vec x)^2- \nonumber\\
 &&\hskip 110pt
 -\,2r_g\Big(\ln kb-\sum_{\ell=2}^\infty
\frac{(2\ell-2)!!}{\ell! \,b^\ell} \sqrt{t^{+2}_\ell +t^{\times2}_\ell}\cos[\ell(\phi_\xi-\phi_\ell)] \Big)\Big)\Big].
  \label{eq:amp-A+}
\end{eqnarray}
As we can see, $A(\vec x)$ is a rather complicated function of the impact parameter $\vec b(b,\phi_\xi)$.  In general, this integral must be evaluated numerically. However, there are two important observations:
\begin{inparaenum}[1)]
\item Even for complicated mass distributions, the contribution of the higher multipole moments of order $\ell$ is suppressed by $1/b^\ell$. In fact, at some distance from the lens, its lensing potential becomes indistinguishable from that of a monopole.
\item In the case of a weakly aspherical lens, multipole moments  are small, making it possible to evaluate (\ref{eq:amp-A+}) using the method of stationary phase with respect to the radial variable, $b$, as we did in \cite{Turyshev-Toth:2021-multipoles,Turyshev-Toth:2021-all-regions}.
\end{inparaenum}

Accordingly, if we evaluate the radial integral in (\ref{eq:amp-A+}) using the method of stationary phase (see \cite{Turyshev-Toth:2017,Turyshev-Toth:2021-multipoles,Turyshev-Toth:2021-all-regions}), the amplification factor (\ref{eq:amp-A+}) takes the form
{}
\begin{eqnarray}
A(\vec x) &=&
\sqrt{2\pi kr_g}e^{i\sigma_0}e^{ik(r_0+r+r_g\ln 4k^2rr_0)}B(\vec x),
  \label{eq:amp-A2-STF}
\end{eqnarray}
where  $B(\vec x)$ is the  generalized complex amplitude of the EM field in case of an arbitrary, weakly aspherical lens:
{}
\begin{eqnarray}
B(\vec x) &=&
\frac{1}{2\pi}\int_0^{2\pi} d\phi_\xi \exp\Big[-ik\Big(\sqrt{\frac{2r_g}{\tilde r}}\rho\cos(\phi_\xi-\phi)-
2r_g\sum_{\ell=2}^\infty
\frac{(2\ell-2)!!}{\ell! \,(\sqrt{2r_g\tilde r})^\ell} \sqrt{t^{+2}_\ell +t^{\times2}_\ell}\cos[\ell(\phi_\xi-\phi_\ell)]\Big)\Big].
  \label{eq:B2-STF}
\end{eqnarray}
Eq.~(\ref{eq:B2-STF}) is a new diffraction integral that extends previous wave-theoretical descriptions of gravitational lensing to the case of a lens with arbitrary structure. Note that instead of (\ref{eq:eik-ph2-STF}), we could use (\ref{eq:eik-ph2-gen}) to obtain an expression identical  to (\ref{eq:B2-STF}) but expressed in terms of spherical harmonics coefficients. The results are equivalent.

The corresponding PSF of the generic lens \cite{Turyshev-Toth:2021-multipoles}, similarly to (\ref{eq:psf}),  is given as
{}
\begin{eqnarray}
{\rm PSF}({\vec x})&=&|B({\vec x})|^2.
 \label{eq:psf=}
\end{eqnarray}
This PSF can be used for the practical modeling of gravitational lenses, especially for imaging of faint sources \cite{Turyshev-Toth:2021-imaging}.

Finally, to study imaging with an extended lens, we must consider the EM field as it is seen through an imaging telescope. To do this, we treat the imaging telescope as a thin lens and perform a Fourier transform of the EM field (\ref{eq:DB-sol-rho}) characterized by the complex amplitude $B(\vec x)$, as given by  (\ref{eq:B2-ext}).  For that, we introduce  ${\vec x}_i$, representing a point on the focal plane of the optical telescope:
{}
\begin{eqnarray}
 \{{\vec x}_i\}&\equiv& (x_i,y_i,0)=\rho_i\big(\cos\phi_i,\sin\phi_i,0\big).
 \label{eq:coord}
\end{eqnarray}

Following \cite{Turyshev-Toth:2021-imaging},  we use the expression for the Poynting vector carried by an EM wave in a vacuum in a flat spacetime and observed on the focal plane of an imaging telescope, ${\vec S}_0({\vec x},{\vec x}_i)$. Then, we obtain the  amplification factor, $\mu({\vec x},{\vec x}_i)$  of the optical system consisting of the SGL and the imaging telescope (i.e., the convolution of the PSF of the SGL with that of an optical telescope), that in the  case of a generic axisymmetric lens takes the form
 {}
\begin{eqnarray}
I({\vec x},{\vec x}_i)=|{\cal A}({\vec x},{\vec x}_i)|^2,
  \label{eq:Pv}
\end{eqnarray}
where $I({\vec x},{\vec x}_i)$ is the intensity distribution corresponding to the image of a point source as seen by the imaging telescope (see details in \cite{Turyshev-Toth:2021-imaging}) and  ${\cal A}({\vec x},{\vec x}_i)$ is the normalized Fourier transform of the complex amplitude $B(\vec x)$ from  (\ref{eq:B2-STF}):
{}
\begin{eqnarray}
{\cal A}({\vec x},{\vec x}_i)&=&
\frac{1}{2\pi}\int_0^{2\pi} d\phi_\xi \,
  \Big(\frac{
2J_1(u(\phi_\xi,\phi_i)\frac{1}{2}d)}{u(\phi_\xi,\phi_i) \frac{1}{2}d}\Big)\times\nonumber\\
&&\hskip 30pt
\times \,
\exp\Big[-ik\Big(\sqrt{\frac{2r_g}{ r}} \rho\cos(\phi_\xi-\phi)-
2r_g\sum_{\ell=2}^\infty
\frac{(2\ell-2)!!}{\ell! \,(\sqrt{2r_g\tilde r})^\ell} \sqrt{t^{+2}_\ell +t^{\times2}_\ell}\cos[\ell(\phi_\xi-\phi_\ell)]\Big)\Big],~~~
  \label{eq:BinscER+}
\end{eqnarray}
with $d$ being the telescope's aperture and $u(\phi_\xi,\phi_i)$ is given by
{}
\begin{eqnarray}
u(\phi_\xi,\phi_i)=\sqrt{\alpha^2+2\alpha\eta_i\cos\big(\phi_\xi-\phi_i\big)+\eta_i^2}, \qquad {\rm where}\qquad \alpha=k\sqrt\frac{2r_g}{r}, \qquad \eta_i=k\frac{\rho_i}{f},
  \label{eq:eps}
\end{eqnarray}
with $\alpha$ and $\eta_i$ characterizing the spatial frequencies of the SGL PSF and its caustic region, while $f$ is the imaging telescope's focal length.

Using the intensity of light observed in the image plane, $I({\vec x},{\vec x}_i)$,  given by (\ref{eq:Pv}) with ${\cal A}({\vec x},{\vec x}_i)$ from  (\ref{eq:BinscER+}) we can study imaging with an extended gravitational lens that has a generic internal structure and mass distribution.

Expressions (\ref{eq:psf=})  and (\ref{eq:Pv})  are the PSF of the extended lens and the intensity of light observed at the image sensor of an imaging telescope. Their optical properties are guided by  (\ref{eq:B2-STF}) and (\ref{eq:BinscER+}), correspondingly. The structure of these expressions is similar  to those studied in \cite{Turyshev-Toth:2021-multipoles,Turyshev-Toth:2021-imaging,Turyshev-Toth:2021-all-regions} where we studied lensing by axisymmetric mass distributions. The primary difference is the fact that expressions (\ref{eq:B2-STF}) and (\ref{eq:BinscER+}) do not have the same rotational axis at each order. In fact, the angle $\phi_\ell$ is different for each STF order, as shown in (\ref{eq:tt52a7}) or (\ref{eq:tt52}), and is set by the unique set of the TT-projected STF multipole mass moments. Based on our prior research \cite{Turyshev-Toth:2021-multipoles}, we know that at each order $\ell$ the PSF will exhibit a unique caustic \cite{Turyshev-Toth:2021-caustics} with the cusps yielding bright images to be observed by the telescope \cite{Turyshev-Toth:2021-imaging}.  This result allows for physically-consistent modeling of realistic gravitational lenses.

\section{Discussion and Conclusions}
\label{sec:end}

In this paper, we studied the optical properties of gravitational lenses with generic mass distributions. To characterize the static external gravitational field produced by such objects, we used the STF tensor representation of the mass multipole moments. Through this route, we were able to obtain the most general solution for the gravitational phase shift, which is the key concept used to characterize the diffraction of light in a gravitational field. This allowed us to develop a wave-optical treatment of gravitational lenses with arbitrary structure and internal mass distribution. In Ref.~\cite{Turyshev-Toth:2021-platonic-STF} we considered the STF multipole moments for several well-known solids possessing uniform density distribution and show how they may be practically computed and used to study gravitational lensing phenomena.

Our results were obtained taking into account that in gravitational lensing, the Schwarzschild radius of the lens, $r_g$, is many orders of magnitude smaller than the typical distances involved, $r_g\ll r,r_0$. Furthermore, as was shown in \cite{Turyshev-Toth:2017,Turyshev-Toth:2020-extend,Turyshev-Toth:2021-multipoles}, the highest amplification is achieved in the case when the light ray's impact parameter, $b$, is of the order of the physical dimensions of a compact lensing object, $R$, that is also much smaller than the typical distances, $R\ll r,r_0$. The resulting conditions on the impact parameter, $b/r\ll1$ and also $b/r_0\ll 1$, make our approximation well-justified for astrophysical lenses. As a result, we demonstrated that the STF form is very useful for practical lensing systems.

The Cartesian STF tensor mass multipole moments are very useful. Although the information about a mass distribution that they posses is identical to that captured by spherical harmonics, the STF representation of a gravitational potential offers technical advantages for problems dealing with light  propagation in a weak gravitational field.  These advantages relate to the fact that the gravitational eikonal equation may now be integrated to any desired order.

We found that at each STF order the gravitational phase shift is fully described by the components of the TT-projected STF tensor moments, as shown by (\ref{eq:tt23*}), which depends only on two parameters $t^+_\ell$ and $t^\times_\ell$, as given by (\ref{eq:eik-ph2-STF}) with (\ref{eq:tt52a7}). These combinations of the STF moments are directly related to the $C_{\ell\ell}$  and $S_{\ell\ell}$ spherical harmonics, as in (\ref{eq:eik-ph2-gen}) with (\ref{eq:tt52}). If either $\{t^+_\ell,t^\times_\ell\}$ or $\{C_{\ell\ell},S_{\ell\ell}\}$  are known or otherwise determined from observations, computation of the PSF is straightforward and is given by a double integral (\ref{eq:amp-A+}). Although this integral expression can be rapidly oscillating, there are regimes where semi-analytical treatments are possible.  Such conditions are realized in the case of impact parameters significantly larger than the physical extent of the lens or when the lens may be treated as weakly aspherical. In these cases, the contributions of the multipole mass moments diminish as $1/b^\ell$ allowing for an iterative solution with the method of stationary phase. As a result, (\ref{eq:amp-A+})  may be reduced to a single integral (\ref{eq:B2-STF}) with finite integration limits.  Similarly, the intensity of light seen at the image sensor of an imaging telescope is govern by a similar integral, as in (\ref{eq:BinscER+}). The resulting integral is manageable with a modest computational cost.

It is quite remarkable that at each multipolar order, only two parameters are required to describe the effect of an extended lens. This is simpler than expected and it applies even to objects without any symmetries.  Although, it is common to account only for the sky plane components of the lensing potential (hence the use of the plane lens approximation, e.g., \cite{Schneider-Ehlers-Falco:1992}), we were able to derive this conclusion rigorously from the first principles.  One consequence is that observations from a single vantage point are limited only to two combinations of the TT-projected STF tensor moments of a gravity field, thus precluding  reconstruction of the full 3-dimensional structure of the mass distribution that is the source of that field (see discussion in \cite{Turyshev-Toth:2021-platonic-STF}). To overcome this limitation, the observing position needs to change. The larger the separation between observing positions, the higher the sensitivity to the effects due to the 3-dimensional structure of an extended mass distribution. We consider this conclusion to be quite fundamental for astronomy.

The new approach presented here is applicable to all weakly aspherical, compact lenses in all lensing regimes (one can demonstrate this by using the approach shown in \cite{Turyshev-Toth:2021-all-regions}). Advantages of our approach arise from the fact that we can now describe lensing by any such  mass distribution using physically justified lens models. These results are new and may be used to study gravitational lensing with a wide range of realistic astrophysical systems, including the SGL.

\begin{acknowledgments}
This work in part was performed at the Jet Propulsion Laboratory, California Institute of Technology, under a contract with the National Aeronautics and Space Administration.
VTT acknowledges the generous support of Plamen Vasilev and other Patreon patrons.
\end{acknowledgments}


\appendix

\section{Coordinate combinations for the lowest STF moments}
\label{sec:stf-comb}

We define the moments a usual:
{}
\begin{eqnarray}
M^{L}=\int d^3{\vec r}' \, \rho({\vec r}')\, x'^L, \qquad {\rm where} \qquad L\in [1,\ell].
\label{eq:Iab}
\end{eqnarray}

The coordinate combinations needed to compute the lowest Cartesian STF multipole moments \cite{Soffel-Han:2019} are:
{}
\begin{eqnarray}
r^2\hat n_{ab}&=&{\rm STF}_{ab}\Big( x^a x^b\Big)=
x^a x^b-\frac{1}{3}r^2\delta^{ab},
\label{eq:sft2}\\
r^3\hat n_{abc}&=&{\rm STF}_{abc}\Big( x^a x^b x^c\Big)=
x^a x^b x^c-\frac{1}{5}r^2\Big(\delta^{ab}x^c+\delta^{bc}x^a+\delta^{ca}x^b\Big),
\label{eq:sft3}\\
r^4\hat n_{abcd}&=&{\rm STF}_{abcd}\Big( x^a x^bx^cx^d\Big)=
x^ax^bx^cx^d-\nonumber\\
&&\hskip -40pt -\,
 \frac{r^2}{7}\Big(x^ax^b\delta^{cd}+x^ax^c\delta^{bd}+x^ax^d\delta^{bc}+x^bx^c\delta^{ad}+x^bx^d\delta^{ac}+x^cx^d\delta^{ab}\Big)+
\frac{r^4}{35}\Big(\delta^{ab}\delta^{cd}+\delta^{ac}\delta^{bd}+\delta^{ad}\delta^{bc}\Big).
\label{eq:sft4}
\end{eqnarray}

\section{Correspondence between the STF mass moments and spherical harmonics}
\label{sec:stf-sph-harm}

To derive the relations between the Cartesian and spherical quadrupole ($\ell = 2$) moments explicitly, we can express the spherical harmonics in terms of Cartesian coordinates. For that we use (\ref{eq:pot_w_0sh}) and write
{}
\begin{eqnarray}
U^{[2]}(\vec r) &=&  \frac{G}{r^3} \Big(P_{20}C_{20} + P_{21}(C_{21} \cos \phi + S_{21}\sin \phi) + P_{22}(C_{22} \cos 2\phi + S_{22} \sin 2\phi)\Big).
\label{eq:U2-sphar}
\end{eqnarray}
Using a spherical coordinate system $ (x = r \sin \theta \cos \phi, y = r \sin \theta \sin \phi, z = r \cos \phi)$, we have
$r^2P_{20} = (2z^2 -x^2 -y^2)/2$,
$r^2P_{21} \cos \phi = 3xz$,
$r^2P_{21} \sin \phi = 3yz$,
$r^2P_{22} \cos 2\phi = 3(x^2 - y^2)$,
$r^2P_{22} \sin 2\phi = 6xy$. Substituting these expressions in (\ref{eq:U2-sphar}), we get
\begin{eqnarray}
U^{[2]}(\vec r) &=&  \frac{GM}{r^5} \Big(C_{20} {\textstyle\frac{1}{2}}(2z^2 - x^2- y^2) + 3C_{21} xz+ 3S_{21}yz + 3C_{22}(x^2-y^2)+6S_{22} xy\Big).
\label{eq:U2-sphar2}
\end{eqnarray}

From (\ref{eq:pot_w_0STF2}), we have the same quantity expressed via the components of the STF quadrupole moment ${\cal T}_{<ab>}$:
{}
\begin{eqnarray}
U^{[2]}(\vec r)&=&GM \frac{3{\cal T}_{<ab>}}{2r^5}x^ax^b=
GM \frac{3}{2r^5}\Big({\cal T}_{11}x^2+2{\cal T}_{12}xy+2{\cal T}_{13}xz+2{\cal T}_{23}yz+{\cal T}_{22}y^2+{\cal T}_{33}z^2\Big).~~~
\label{eq:pot_w_0STF2*}
\end{eqnarray}

Equating the terms with the same powers of $x,y,z$  between the from of the potential in terms of spherical harmonics present in (\ref{eq:U2-sphar2}) and that expressed via the STF mass quadrupole in (\ref{eq:pot_w_0STF2*}) yields the following relations:
{}
\begin{eqnarray}
{\cal T}_{11}&=&\Big(-{\textstyle\frac{1}{3}}C_{20}+2C_{22}\Big)R^2, \qquad  {\cal T}_{12}=2S_{22}R^2, \nonumber\\
{\cal T}_{22}&=&\Big(-{\textstyle\frac{1}{3}}C_{20}-2C_{22}\Big)R^2, \qquad {\cal T}_{13}=-C_{21}R^2,\nonumber\\
{\cal T}_{33}&=&{\textstyle\frac{2}{3}} C_{20} R^2, \qquad\qquad \qquad\quad~~
{\cal T}_{23}=-S_{21}R^2.
\label{eq:sp-harm2*}
\end{eqnarray}

Following the same approach, we can establish the corresponding relationships between the octupole and hexadecapole moments and the appropriate  spherical harmonics coefficients given by (\ref{eq:sp-harm3}) and (\ref{eq:sp-harm4}).

\section{Computing particular cases}
\label{sec:cases}

To demonstrate the practical utility of our results, we derive several low order terms in (\ref{eq:eik-ph2}). First, we recognize that with $\vec k$ being constant, the two-dimensional vector $\vec b$ and the one-dimensional quantity $\tau$, from (\ref{eq:x-Newt*=0})--(\ref{eq:b0}), may be treated as two independent variables, yielding $dx^a=db^a+k^ad\tau$ and also ${\partial }/{\partial x^a}={\partial}/{\partial b^a}+k^a {\partial}/{\partial \tau}$, where differentiation with respect to $\vec{b}$ is carried out in two dimensions only, which is indicated by the hatted notation. Then, to compute the needed partial derivatives in (\ref{eq:eik-ph2}), with respect to the vector of the impact parameter, $\hat\partial_a\equiv \partial/\partial b^a\equiv(\partial/\partial b^x,\partial/\partial b^y,0)$ in our chosen Cartesian coordinate system, we may formally write
{}
\begin{eqnarray}
\frac{\partial x^a}{\partial x^b}= \delta^a_b= \Big\{{\hat\partial}_b+k_b \frac{\partial}{\partial \tau}\Big\}\Big\{b^a+k^a \tau +{\cal O}(r_g)\Big\}=\hat\partial_b b^a+k^a k_b+{\cal O}(r_g).
\label{eq:eik-exp+}
\end{eqnarray}
By re-arranging the terms in this identity, we obtain the following useful  expression (see also \cite{Kopeikin:1997,Soffel-Han:2019}):
{}
\begin{eqnarray}
\hat\partial_b b^a&=&
\delta^a_b -k^a k_b.
\label{eq:eik-exp}
\end{eqnarray}
This result essentially is the projection operator onto the plane perpendicular to $\vec k$, namely $P^{ab}=\delta^{ab}-k^ak^b$; this plane, given the parametrization (\ref{eq:note-k})--(\ref{eq:note-b}), is the plane of the impact parameter $\vec b$.

With result (\ref{eq:eik-exp}), we compute the partial derivatives present in (\ref{eq:eik-ph2}) for several low order terms, namely $\ell=2,3,4$:
{}
\begin{eqnarray}
\hat\partial_a \ln kb&=&  \frac{b_a}{b^2}, \\
\hat\partial^2_{ab} \ln kb&=&  \frac{1}{b^2}\Big(\delta_{ab}-k_ak_b\Big)-\frac{2b_ab_b}{b^4},
\label{eq:dab2}\\
\hat\partial^3_{abc} \ln kb&=& - \frac{2b_a}{b^4}\Big(\delta_{bc}-k_bk_c\Big)-\frac{2b_b}{b^4}\Big(\delta_{ac}-k_ak_c\Big)- \frac{2b_c}{b^4}\Big(\delta_{ab}-k_ak_b\Big)+\frac{8b_ab_bb_c}{b^6},
\label{eq:dab3}\\
\hat\partial^4_{abcd} \ln kb&=&  - \frac{2}{b^4}\Big(\delta_{bc}-k_bk_c\Big)\Big(\delta_{ad}-k_ak_d\Big)-\frac{2}{b^4}\Big(\delta_{ac}-k_ak_c\Big)\Big(\delta_{bd}-k_bk_d\Big)-\frac{2}{b^4}\Big(\delta_{ab}-k_ak_b\Big)\Big(\delta_{cd}-k_ck_d\Big)+\nonumber\\
&&\hskip 0pt+\,\frac{8}{b^6}\Big(b_ab_b\big(\delta_{cd}-k_ck_d\big)+b_ab_c\big(\delta_{bd}-k_bk_d\big)+b_ab_d\big(\delta_{bc}-k_bk_c\big)+b_bb_c\big(\delta_{ad}-k_ak_d\big)+\nonumber\\
&&\hskip 16pt+\, b_bb_d\big(\delta_{ac}-k_ak_c\big)+b_cb_d\big(\delta_{ab}-k_ak_b\big)\Big)-\frac{48b_ab_bb_cb_d}{b^8},
\label{eq:dab4}
\end{eqnarray}
where $ b^a$ is the  $a$-th component of the vector impact parameter.

\section{Introducing projection operators}
\label{sec:projops}

Recognizing the fact that (\ref{eq:eik-exp}) is the projection operator:
\begin{eqnarray}
P^{ab}&=& \delta^{ab} -k^a k^b,
\label{eq:eik-exp1}
\end{eqnarray}
which projects on the transverse direction with respect to $\vec k$ or on the plane of the impact parameter $\vec b$, we present results (\ref{eq:dab2})--(\ref{eq:dab4}) in the following identical form:
{}
\begin{eqnarray}
\hat\partial_a \ln kb&=&
P_{ai}\frac{b^i}{b^2}, \\
\hat\partial^2_{ab} \ln kb&=&
-\Big(P_{ai}P_{bj}-{\textstyle\frac{1}{2}}P_{ab}P_{ij}\Big)\frac{1}{b^2}\Big(2m^im^j-\delta^{ij}\Big),
\label{eq:dab21}\\
\hat\partial^3_{abc} \ln kb&=&
\Big(P_{ai}P_{bj}P_{ck}-{\textstyle\frac{1}{4}}P_{ab}P_{ci}P_{jk}-{\textstyle\frac{1}{4}}P_{ac}P_{bi}P_{jk}-{\textstyle\frac{1}{4}}P_{bc}P_{ai}P_{jk}\Big)
\frac{2}{b^3}\Big(4m^im^jm^k- \delta^{ij}m^k-\delta^{ik}m^j- \delta^{kj}m^i\Big),~~~~~
\label{eq:dab31}\\
\hat\partial^4_{abcd} \ln kb&=&
-\bigg\{P_{ai}P_{bj}P_{ck}P_{dl}-
{\textstyle\frac{1}{6}}\Big(
P_{ab}P_{ij}P_{ck}P_{dl}+P_{ac}P_{ik}P_{bj}P_{dl}+P_{ad}P_{il}P_{bj}P_{ck}+P_{bc}P_{jk}P_{ai}P_{dl}+
\nonumber\\
&&\hskip 0pt+\,
P_{bd}P_{jl}P_{ai}P_{ck}+P_{cd}P_{kl}P_{ai}P_{bj}\Big)+
{\textstyle\frac{1}{24}}\Big(
P_{ab}P_{cd}+P_{ac}P_{bd}+P_{bc}P_{ad}\Big)P_{ij}P_{kl}
\bigg\}
\frac{2}{b^4}\bigg\{24m^im^jm^km^l-
\nonumber\\
&&\hskip -30pt-\,
4\Big(\delta^{il}m^jm^k+\delta^{jl}m^im^k+\delta^{kl}m^im^j+\delta^{ij}m^km^l+\delta^{ik}m^jm^l+\delta^{kj}m^im^l\Big)+
\Big(\delta^{ij}\delta^{kl}+\delta^{ik}\delta^{jl}+\delta^{il}\delta^{kj}\Big)\bigg\},
\label{eq:dab41}
\end{eqnarray}
where $ b^a=b m^a$ is the  $a$-th component of the vector of the impact parameter, with $b$ being its magnitude and $m^a$ its unit vector, see definitions given in parametrization (\ref{eq:note-k})--(\ref{eq:note-b}).

With the results (\ref{eq:dab21})--(\ref{eq:dab41}), we express the appropriate terms of $ {\cal T}^{<a_1...a_\ell>} \hat\partial_{<a_1...a_\ell>}\ln kb$, as
{}
\begin{eqnarray}
 {\cal T}^{<ab>} \hat\partial_{<ab>}\ln kb&=&- {\cal T}^{<ab>} \Big(P_{ai}P_{bj}-{\textstyle\frac{1}{2}}P_{ab}P_{ij}\Big)\frac{1}{b^2}\Big(2m^im^j-\delta^{ij}\Big)=\nonumber\\
 &=&-\big[{\cal T}^{<ij>}\big]^{\rm TT}\frac{1}{b^2}\Big(2m^im^j-\delta^{ij}\Big)\equiv -\big[{\cal T}^{<ij>}\big]^{\rm TT}\frac{2}{b^2}\,m^im^j,
  \label{eq:tt2}\\
  {\cal T}^{<abc>} \hat\partial_{<abc>}\ln kb&=&{\cal T}^{<abc>} \Big(P_{ai}P_{bj}P_{ck}-{\textstyle\frac{1}{4}}P_{ab}P_{ci}P_{jk}-{\textstyle\frac{1}{4}}P_{ac}P_{bi}P_{jk}-{\textstyle\frac{1}{4}}P_{bc}P_{ai}P_{jk}\Big)\times\nonumber\\
  &&
  \hskip35pt
  \times\, \frac{2}{b^3}\Big(4m^im^jm^k- \delta^{ij}m^k-\delta^{ik}m^j- \delta^{kj}m^i\Big)=\nonumber\\
  &&\hskip-40pt
  =\,\big[{\cal T}^{<ijk>}\big]^{\rm TT}\frac{2}{b^3}\Big(4m^im^jm^k- \delta^{ij}m^k-\delta^{ik}m^j- \delta^{kj}m^i\Big)\equiv
  \big[{\cal T}^{<ijk>}\big]^{\rm TT}\frac{8}{b^3}\,m^im^jm^k,
  \label{eq:tt3} \\
   {\cal T}^{<abcd>} \hat\partial_{<abcd>}\ln kb&=&-{\cal T}^{<abcd>} \bigg\{P_{ai}P_{bj}P_{ck}P_{dl}-
{\textstyle\frac{1}{6}}\Big(
P_{ab}P_{ij}P_{ck}P_{dl}+P_{ac}P_{ik}P_{bj}P_{dl}+P_{ad}P_{il}P_{bj}P_{ck}+\nonumber\\
&&\hskip -105pt+\,
P_{bc}P_{jk}P_{ai}P_{dl}+P_{bd}P_{jl}P_{ai}P_{ck}+P_{cd}P_{kl}P_{ai}P_{bj}\Big)+
{\textstyle\frac{1}{24}}\Big(
P_{ab}P_{cd}+P_{ac}P_{bd}+P_{bc}P_{ad}\Big)P_{ij}P_{kl}
\bigg\}\frac{2}{b^4}\bigg\{24m^im^jm^km^l-\nonumber\\
&&
\hskip-70pt
-4\Big(\delta^{il}m^jm^k+\delta^{jl}m^im^k+\delta^{kl}m^im^j+\delta^{ij}m^km^l+\delta^{ik}m^jm^l+\delta^{kj}m^im^l\Big)+
\delta^{ij}\delta^{kl}+\delta^{ik}\delta^{jl}+\delta^{il}\delta^{kj}\bigg\}=\nonumber\\
&&
\hskip-95pt
=\,-
\big[{\cal T}^{<ijkl>}\big]^{\rm TT}\frac{2}{b^4}\bigg\{24m^im^jm^km^l-
4\Big(\delta^{il}m^jm^k+\delta^{jl}m^im^k+\delta^{kl}m^im^j+\delta^{ij}m^km^l+\delta^{ik}m^jm^l+\delta^{kj}m^im^l\Big)+
\nonumber\\
&&\hskip-40pt +\,
\delta^{ij}\delta^{kl}+ \delta^{ik}\delta^{jl}+\delta^{il}\delta^{kj}\bigg\}=
-\big[{\cal T}^{<ijkl>}\big]^{\rm TT}\frac{48}{b^4}\,m^im^jm^km^l,
  \label{eq:tt4}
\end{eqnarray}
where the superscript {\tt TT} stands for transverse traceless projection in the lens plane.

As a result, we have the following expressions:
{}
\begin{eqnarray}
 {\cal T}^{<ab>} \hat\partial_{<ab>}\ln kb&=& -\big[{\cal T}^{<ij>}\big]^{\rm TT}\frac{2}{b^2}\,m^im^j,
  \label{eq:tt23*f}\\
  {\cal T}^{<abc>} \hat\partial_{<abc>}\ln kb&=&
  \big[{\cal T}^{<ijk>}\big]^{\rm TT}\frac{8}{b^3}\,m^im^jm^k,
  \label{eq:tt33*f} \\
   {\cal T}^{<abcd>} \hat\partial_{<abcd>}\ln kb&=&-
\big[{\cal T}^{<ijkl>}\big]^{\rm TT}\frac{48}{b^4}\,m^im^jm^km^l.
  \label{eq:tt43*f}
\end{eqnarray}
Results, similar to (\ref{eq:tt23*f})--(\ref{eq:tt43*f}), were verified through $\ell=10$, thus enabling a generalization to an arbitrary $\ell$.
\section{Polarization matrices}
\label{sec:pol-mat}

We computed the explicit forms of the polarization matrices $U^+_{a_1...a_\ell}\equiv U^+_\ell$ and $U^\times_{a_1...a_\ell}\equiv U^\times_\ell$ entering expression (\ref{eq:tt5*a}) to several orders, $\ell \in\{2,3,4\}$:
{}
\begin{eqnarray}
U^+_{2}=\begin{bmatrix}
1 & 0 & 0\\
0 & -1 & 0 \\
0 & 0 & 0
\end{bmatrix}, \qquad
U^\times_{2}=\begin{bmatrix}
0 & 1 & 0\\
1 & 0 & 0 \\
0 & 0 & 0
\end{bmatrix},
\label{eq:U2}
\end{eqnarray}

\begin{eqnarray}
U^+_{3}=\begin{bmatrix}
\begin{pmatrix}
1\\
0\\
0
\end{pmatrix} &
\begin{pmatrix}
0\\
-1\\
0
\end{pmatrix}
&
\begin{pmatrix}
0\\
0\\
0
\end{pmatrix}\\
\begin{pmatrix}
0\\
-1\\
0
\end{pmatrix} &
\begin{pmatrix}
-1\\
0\\
0
\end{pmatrix}&
\begin{pmatrix}
0\\
0\\
0
\end{pmatrix} \\
\begin{pmatrix}
0\\
0\\
0
\end{pmatrix} &
\begin{pmatrix}
0\\
0\\
0
\end{pmatrix} &
\begin{pmatrix}
0\\
0\\
0
\end{pmatrix}
\end{bmatrix}, \qquad
U^\times_{3}=\begin{bmatrix}
\begin{pmatrix}
0\\
1\\
0
\end{pmatrix} &
\begin{pmatrix}
1\\
0\\
0
\end{pmatrix}
&
\begin{pmatrix}
0\\
0\\
0
\end{pmatrix}\\
\begin{pmatrix}
1\\
0\\
0
\end{pmatrix} &
\begin{pmatrix}
0\\
-1\\
0
\end{pmatrix}&
\begin{pmatrix}
0\\
0\\
0
\end{pmatrix} \\
\begin{pmatrix}
0\\
0\\
0
\end{pmatrix} &
\begin{pmatrix}
0\\
0\\
0
\end{pmatrix} &
\begin{pmatrix}
0\\
0\\
0
\end{pmatrix}
\end{bmatrix},
\label{eq:U3}
\end{eqnarray}

\begin{eqnarray}
U^+_4=\begin{bmatrix}
\begin{pmatrix}
1 & 0 & 0\\
0 &-1 & 0\\
0 & 0 & 0
\end{pmatrix} &
\begin{pmatrix}
0 & -1 & 0\\
-1 & 0 & 0\\
0 & 0 & 0
\end{pmatrix}
&
\begin{pmatrix}
0 & 0 & 0\\
0 & 0 & 0\\
0 & 0 & 0
\end{pmatrix}\\
\begin{pmatrix}
0 & -1 & 0\\
-1  & 0 & 0\\
0 & 0 & 0
\end{pmatrix} &
\begin{pmatrix}
-1 & 0 & 0\\
0 & 1 & 0\\
0 & 0 & 0
\end{pmatrix}&
\begin{pmatrix}
0 & 0 & 0\\
0 & 0 & 0\\\
0 & 0 & 0\
\end{pmatrix} \\
\begin{pmatrix}
0 & 0 & 0\\\
0 & 0 & 0\\\
0 & 0 & 0\
\end{pmatrix} &
\begin{pmatrix}
0 & 0 & 0\\\
0 & 0 & 0\\\
0 & 0 & 0\
\end{pmatrix} &
\begin{pmatrix}
0 & 0 & 0\\\
0 & 0 & 0\\\
0 & 0 & 0\
\end{pmatrix}
\end{bmatrix}, \qquad
U^\times_4=\begin{bmatrix}
\begin{pmatrix}
0 & 1 & 0\\
1 &0 & 0\\
0 & 0 & 0
\end{pmatrix} &
\begin{pmatrix}
1 & 0 & 0\\
0 & -1 & 0\\
0 & 0 & 0
\end{pmatrix}
&
\begin{pmatrix}
0 & 0 & 0\\
0 & 0 & 0\\
0 & 0 & 0
\end{pmatrix}\\
\begin{pmatrix}
1 & 0 & 0\\
0  & -1 & 0\\
0 & 0 & 0
\end{pmatrix} &
\begin{pmatrix}
0 & -1 & 0\\
-1 & 0 & 0\\
0 & 0 & 0
\end{pmatrix}&
\begin{pmatrix}
0 & 0 & 0\\
0 & 0 & 0\\\
0 & 0 & 0\
\end{pmatrix} \\
\begin{pmatrix}
0 & 0 & 0\\\
0 & 0 & 0\\\
0 & 0 & 0\
\end{pmatrix} &
\begin{pmatrix}
0 & 0 & 0\\\
0 & 0 & 0\\\
0 & 0 & 0\
\end{pmatrix} &
\begin{pmatrix}
0 & 0 & 0\\\
0 & 0 & 0\\\
0 & 0 & 0\
\end{pmatrix}
\end{bmatrix}.
\label{eq:U4}
\end{eqnarray}

One can see that the basis vectors  $ U^+_\ell$ and $U^\times_\ell$  are block matrices where the structure at each consecutive order replicates the structure at $\ell=2$ with different signs. Similar structures, but with ever increasing complexity, are evident for each order beyond those shown by expressions (\ref{eq:U2})--(\ref{eq:U4}).  Each block within the matrices is responsible for a single rotation, that being combined with others leads to a rotation by $\ell\phi_\ell$, in analogy to the Chebychev polynomials.

Using expressions (\ref{eq:U2})--(\ref{eq:U4}), we can verify that the following relations exist  for $\ell \in\{2,3,4\}$:
{}
\begin{eqnarray}
U^+_{a_1...a_\ell}\,m^{a_1...}m^{a_\ell}=\cos[\ell\phi_\xi],
\qquad
U^\times_{a_1...a_\ell} \,m^{a_1...}m^{a_\ell}=\sin[\ell\phi_\xi], \qquad
  \label{eq:tt52q*}
\end{eqnarray}
where the components $m^a$ of the vector impact parameter are given by (\ref{eq:note-b}). In fact, using computer algebra, we were able to verify these relationships up to $\ell=10$, which may lead to generalization of an arbitrary order  $\ell$.

\section{Large impact parameter approximation}
\label{sec:beyond}

We recall that (\ref{eq:eik-ph-shift}) was obtained from  (\ref{eq:delta-eik-gen}) by assuming $b/r_0\ll1$ and $b/r\ll1$. As a result, we are not sensitive to the longitudinal components of the gravitational field as expressed via the STF multipole moments. However, in the case when $b/r_0\sim1$ and $b/r\sim1$, the phase shift (\ref{eq:delta-eik-gen}) will contain all STF components. To demonstrate this, we compute the phase shift for the quadrupole moment, $\ell=2$. The derivatives present in (\ref{eq:delta-eik-gen}) are
{}
\begin{eqnarray}
\hat\partial_a \ln k\Big(\sqrt{b^2+\tau^2}+\tau\Big)&=& \frac{1}{\sqrt{b^2+\tau^2}+\tau}\frac{P_{ai}b^i}{\sqrt{b^2+\tau^2}},
\label{eq:dab510}\\
\hat\partial^2_{ab} \ln k\Big(\sqrt{b^2+\tau^2}+\tau\Big)&=& \frac{P_{ai}P_{bj}}{\sqrt{b^2+\tau^2}+\tau}\frac{1}{\sqrt{b^2+\tau^2}}\Big\{\delta^{ij}-\frac{1}{\sqrt{b^2+\tau^2}}\Big(\frac{1}{\sqrt{b^2+\tau^2}+\tau}+\frac{1}{\sqrt{b^2+\tau^2}}\Big)b^ib^j\Big\},
\label{eq:dab51}
\end{eqnarray}

We also need the following:
\vskip -12pt
\begin{eqnarray}
  \sum_{p=1}^2 \frac{\ell!}{p!(\ell-p)!}k_{<a_1}...k_{a_p} \hat\partial_{a_{p+1}}... \hat\partial_{a_\ell>}
\frac{\partial^{p-1}}{\partial \tau^{p-1}}  \frac{1}{\sqrt{b^2+\tau^2}}&=&  2k_{<a}\hat\partial_{b>} \frac{1}{\sqrt{b^2+\tau^2}}+k_{<a}k_{b>} \frac{\partial}{\partial \tau} \frac{1}{\sqrt{b^2+\tau^2}}=\nonumber\\
&=&-\frac{1}{(b^2+\tau^2)^\frac{3}{2}}\Big(2(k_ab_b+k_bb_a)+\tau(k_ak_b-{\textstyle\frac{1}{3}}\delta_{ab})\Big).
 \label{eq:dab61}
\end{eqnarray}
Clearly, the same expressions exist for terms that depend on $\tau_0$.

Using the result (\ref{eq:b0}) we determine that the following relations are valid to ${\cal O} (r_g)$:
{}
\begin{eqnarray}
r= \sqrt{b^2+\tau^2}, \qquad r+({\vec k}\cdot{\vec r}) =\sqrt{b^2+\tau^2} +\tau.
 \label{eq:notat}
\end{eqnarray}
Then, using the derivatives (\ref{eq:dab51})--(\ref{eq:dab61}) and relying on the notations (\ref{eq:notat}) from (\ref{eq:delta-D*-av0WKB+1*}) we have the following expression for the gravitational eikonal phase shift induced by the quadrupole STF moment, $\ell=2$:
{}
\begin{eqnarray}
\varphi_2(\vec b)
&=& {\textstyle\frac{1}{2}} kr_g
{\cal T}^{<ab>}
\Big\{
\Big(m_am_b-{\textstyle\frac{1}{3}}\delta_{ab}\Big)\Big(\frac{4}{b^2}-
\frac{b^2\big(2r+({\vec k}\cdot{\vec r})\big)}{r^3\big(r+({\vec k}\cdot{\vec r})\big)^2}-
\frac{b^2\big(2r_0+({\vec k}\cdot{\vec r}_0)\big)}{r^3_0\big(r_0+({\vec k}\cdot{\vec r}_0)\big)^2}\Big)+\nonumber\\
 &&\hskip -65pt +\,
\Big(k_ak_b-{\textstyle\frac{1}{3}}\delta_{ab}\Big)\Big(\frac{2}{b^2}-
\frac{1}{r\big(r+({\vec k}\cdot{\vec r})\big)}-\frac{({\vec k}\cdot{\vec r})}{r^3}-\frac{1}{r_0\big(r_0+({\vec k}\cdot{\vec r}_0)\big)}-\frac{({\vec k}\cdot{\vec r}_0)}{r^3_0}\Big)-
2\Big(k_am_b+k_bm_a\Big)\Big(
\frac{b}{r^3}+\frac{b}{r^3_0}\Big)
\Big\}.
\label{eq:delta-eik-gen*}
\end{eqnarray}

Using the parametrization (\ref{eq:note-k})--(\ref{eq:note-b}), we transform the appropriate terms in (\ref{eq:delta-eik-gen*}) as
{}
\begin{eqnarray}
{\cal T}^{<ab>}
\Big(m_am_b-{\textstyle\frac{1}{3}}\delta_{ab}\Big)&=&{\textstyle\frac{1}{2}}({\cal T}_{11}+{\cal T}_{22})+{\textstyle\frac{1}{2}}({\cal T}_{11}-{\cal T}_{22})\cos2\phi_\xi+{\cal T}_{12}\sin2\phi_\xi,
\label{eq:delta-mm}\\
{\cal T}^{<ab>} \Big(k_ak_b-{\textstyle\frac{1}{3}}\delta_{ab}\Big)&=&
{\cal T}_{33},
\label{eq:delta-kk}\\
{\cal T}^{<ab>}\Big(k_am_b+k_bm_a\Big)&=&2{\cal T}_{13}\cos2\phi_\xi+2{\cal T}_{23}\sin\phi_\xi.
\label{eq:delta-km}
\end{eqnarray}
With these results and using the fact that ${\cal T}^{<ab>}$ is an STF tensor, ${\cal T}_{11}+{\cal T}_{22}+{\cal T}_{33}=0$, we transform (\ref{eq:delta-eik-gen*}) as
{}
\begin{eqnarray}
\varphi_2(\vec b)
&=& {\textstyle\frac{1}{2}} kr_g
\Big\{
\Big({\textstyle\frac{1}{2}}({\cal T}_{11}-{\cal T}_{22})\cos2\phi_\xi+{\cal T}_{12}\sin2\phi_\xi \Big) \Big(\frac{4}{b^2}-
\frac{b^2\big(2r+({\vec k}\cdot{\vec r})\big)}{r^3\big(r+({\vec k}\cdot{\vec r})\big)^2}
-\frac{b^2\big(2r_0+({\vec k}\cdot{\vec r}_0)\big)}{r^3_0\big(r_0+({\vec k}\cdot{\vec r}_0)\big)^2}\Big)-\nonumber\\
 &&\hskip -40pt -\,
{\cal T}_{33}\Big(\frac{1}{r\big(r+({\vec k}\cdot{\vec r})\big)}\Big(1-
\frac{b^2\big(2r+({\vec k}\cdot{\vec r})\big)}{2r^2\big(r+({\vec k}\cdot{\vec r})\big)^2}\Big)+\frac{ ({\vec k}\cdot{\vec r})}{r^3}+\frac{1}{r_0\big(r+({\vec k}\cdot{\vec r}_0)\big)}\Big(1-
\frac{b^2\big(2r_0+({\vec k}\cdot{\vec r}_0)\big)}{2r^2_0\big(r_0+({\vec k}\cdot{\vec r}_0)\big)^2}\Big)+\frac{ ({\vec k}\cdot{\vec r}_0)}{r^3_0}\Big)-\nonumber\\
 &&\hskip 40pt -\,
4\Big({\cal T}_{13}\cos\phi_\xi+{\cal T}_{23}\sin\phi_\xi\Big)
\Big(\frac{b}{r^3}+\frac{b}{r_0^3}\Big)
\Big\}.
\label{eq:delta-eik-gen2*}
\end{eqnarray}
As expected, this expression contains all six components of ${\cal T}^{<ab>}$ with only five of them being independent. In realistic lensing geometries $b/r\ll1$ and $b/r_0\ll 1$, which brings  (\ref{eq:delta-eik-gen2*}) to the form (\ref{eq:eik-phas2q}) that features only the  transverse components of the quadrupole moment.

Using the relationship between ${\cal T}^{<ab>}$ and the spherical harmonic coefficients from (\ref{eq:sp-harm2}), we transform (\ref{eq:delta-eik-gen2*}) as
{}
\begin{eqnarray}
\varphi_2(\vec b)
&=& kr_gR^2
\Big\{
\Big(C_{22}\cos2\phi_\xi+S_{22}\sin2\phi_\xi \Big) \Big(\frac{4}{b^2}-
\frac{b^2\big(2r+({\vec k}\cdot{\vec r})\big)}{r^3\big(r+({\vec k}\cdot{\vec r})\big)^2}
-\frac{b^2\big(2r_0+({\vec k}\cdot{\vec r}_0)\big)}{r^3_0\big(r_0+({\vec k}\cdot{\vec r}_0)\big)^2}\Big)-\nonumber\\
 &&\hskip -40pt -\,
{\textstyle\frac{1}{3}}C_{20} \Big(\frac{1}{r\big(r+({\vec k}\cdot{\vec r})\big)}\Big(1-
\frac{b^2\big(2r+({\vec k}\cdot{\vec r})\big)}{2r^2\big(r+({\vec k}\cdot{\vec r})\big)^2}\Big)+\frac{ ({\vec k}\cdot{\vec r})}{r^3}+\frac{1}{r_0\big(r+({\vec k}\cdot{\vec r}_0)\big)}\Big(1-
\frac{b^2\big(2r_0+({\vec k}\cdot{\vec r}_0)\big)}{2r^2_0\big(r_0+({\vec k}\cdot{\vec r}_0)\big)^2}\Big)+\frac{ ({\vec k}\cdot{\vec r}_0)}{r^3_0}\Big)+\nonumber\\
 &&\hskip 40pt +\,
2\Big(C_{21}\cos\phi_\xi+S_{31}\sin\phi_\xi\Big)
\Big(\frac{b}{r^3}+\frac{b}{r_0^3}\Big)
\Big\}.
\label{eq:delta-eik-gen24*7}
\end{eqnarray}
We recall that  $r=r(b,\tau)$ and $r_0=r(b,\tau_0)$ are given by (\ref{eq:notat}). Thus, in the case when $b/r,r/r_0\ll1$, we recover the earlier result (\ref{eq:eik-phas2q*}), where $\varphi_2(\vec b)=4kr_g (R/b)^2\sqrt{C^2_{22}+S^2_{22}}\cos[2(\phi_\xi-\phi_2)])(1+{\cal O}(b^2/r^2))$.

\end{document}